\documentclass[fleqn,10pt]{wlscirep}

\usepackage[utf8]{inputenc}
\usepackage[T1]{fontenc}

\providecommand{\st}[1]{} % strikeout command
\providecommand{\hl}[1]{#1} % To highlight text using \hl

\usepackage{multirow}

\usepackage[UKenglish]{isodate}% http://ctan.org/pkg/isodate

\usepackage{amsmath,amssymb,amscd,mathtools}
\usepackage{graphicx}
\graphicspath{{./figures/}}

\providecommand{\href}[2]{\texttt{#2}}
\providecommand{\url}[1]{\texttt{#1}}

\newcommand{\bea}{\begin{eqnarray}}
\newcommand{\eea}{\end{eqnarray}}
\newcommand{\beq}{\begin{equation}}
\newcommand{\eeq}{\end{equation}}

\newcommand{\ra}{\rightarrow}

\newcommand{\tdef}[1]{\textit{#1}}

\newcommand{\Author}[1]{#1}
\newcommand{\Journal}[1]{#1}
\newcommand{\Year}[1]{(#1)}
\newcommand{\Volume}[1]{#1}
\newcommand{\Number}[1]{}
\newcommand{\Pages}[1]{#1}

\providecommand{\eqref}[1]{(\ref{#1})}
\newcommand{\figref}[1]{Figure~\ref{#1}}
\newcommand{\Figref}[1]{Figure~\ref{#1}}
\newcommand{\subfigref}[2]{Figure~\ref{#1}#2}
\newcommand{\tabref}[1]{Table~\ref{#1}}
\newcommand{\Tabref}[1]{Table~\ref{#1}}
\newcommand{\secref}[1]{Section~\ref{#1}}

\newcommand{\tsevec}[1]{\mathbf{#1}}
\newcommand{\tsemat}[1]{{\mathbf{\textsf{#1}}}}

\newcommand{\texpect}[1]{\langle #1 \rangle}

\newcommand{\Ecal}{\mathcal{E}}
\newcommand{\Gcal}{\mathcal{G}}

\newcommand{\Mcal}{\mathcal{M}}
\newcommand{\Ncal}{\mathcal{N}}

\newcommand{\Tcal}{\mathcal{T}}
\newcommand{\Vcal}{\mathcal{V}}

\newcommand{\phat}{\hat{p}}
\newcommand{\qhat}{\hat{q}}

\newcommand{\unitmat}{\tsemat{I}}
\newcommand{\Amat}{\tsemat{A}}

\newcommand{\Dmat}{\tsemat{D}}

\newcommand{\Lmat}{\tsemat{L}}

\newcommand{\Pout}{P^\mathrm{(out)}}
\newcommand{\Poutmat}{\tsemat{P}^\mathrm{(out)}}

\newcommand{\Qmat}{\tsemat{Q}}

\newcommand{\That}{\hat{T}}
\newcommand{\Tsf}{\mathsf{T}}
\newcommand{\Tsfhat}{\widehat{\mathsf{T}}}
\newcommand{\Tsfpw}{\Tsf^{(\mathrm{pw})}}
\newcommand{\Tsfhatpw}{\widehat{\mathsf{T}}^{\mathrm{(pw)}}}

\newcommand{\Prob}{\mathrm{Pr}}

\newcommand{\Triad}{\mathsf{M}}
\newcommand{\Triadset}{\Mcal}
\newcommand{\Tripletset}{\Tcal}

\newcommand{\wvec}{\tsevec{w}}

\title{Higher-Order Temporal Network Effects through Triplet Evolution\thanks{{Published as \emph{Scientific Reports} \textbf{11} (2021) 15419. DOI \href{http://doi.org/10.1038/s41598-021-94389-w}{\texttt{10.1038/s41598-021-94389-w}} }}}
\author[1,2c*]{Qing Yao}
\author[1*]{Bingsheng Chen}
\author[1]{\href{http://www.imperial.ac.uk/people/t.evans}{Tim S.\ Evans}}
\author[1]{ \href{http://www.imperial.ac.uk/people/k.christensen}{Kim Christensen}}
\affil[1]{\href{https://www.imperial.ac.uk/physics/}{Blackett Laboratory} and \href{https://www.imperial.ac.uk/complexity-science}{Centre for Complexity Science}, Imperial College London, South Kensington Campus, London SW7 2AZ, United Kingdom}
\affil[2]{School of Systems Science, Beijing Normal University, 100875 Beijing, China}

\affil[c]{corresponding author~q.yao15@imperial.ac.uk}

\affil[*]{these authors contributed equally to this work}

\begin{abstract}
    We study the evolution of networks through `triplets' --- three-node graphlets. We develop a method to compute a transition matrix to describe the evolution of triplets in temporal networks. To identify the importance of higher-order interactions in the evolution of networks, we compare both artificial and real-world data to a model based on pairwise interactions only. The significant differences between the computed matrix and the calculated matrix from the fitted parameters demonstrate that non-pairwise interactions exist for various real-world systems in space and time, such as our data sets. Furthermore, this also reveals that different patterns of higher-order interaction are involved in different real-world situations.

    To test our approach, we then use these transition matrices as the basis of a link prediction algorithm. We investigate our algorithm's performance on four temporal networks, comparing our approach against ten other link prediction methods. Our results show that higher-order interactions in both space and time play a crucial role in the evolution of networks as we find our method, along with two other methods based on non-local interactions, give the best overall performance. The results also confirm the concept that the higher-order interaction patterns, i.e., triplet dynamics, can help us understand and predict the evolution of different real-world systems.
\end{abstract}

\begin{document}

\flushbottom
\maketitle
\thispagestyle{empty}

% \noindent Please note: Abbreviations should be introduced at the first mention in the main text – no abbreviations lists. Suggested structure of main text (not enforced) is provided below.

\section*{Introduction}

The collective behaviour of a complex system cannot be understood and predicted by considering the basic units of the system in isolation \cite{APM1972,BFG2020}. Complex  \st{Networks}\hl{networks}~\cite{N2018, SS2001} are often used to represent complex systems \st{and then}\hl{in which} the basic unit of interaction, the edge, represents pairwise interactions. However, in systems represented by networks there are often local processes, known as higher-order interactions, where groups of more than two participants interact and the set of pairwise interactions does not capture the whole picture~\cite{RMI2019}.
Research has revealed that many empirical systems display higher-order interactions, for example social systems~\cite{R53,G73,BAD2016}, neuroscience~\cite{ERD2011,PGP2014,ELKP19}, biology~\cite{SAD2018}, and ecology~\cite{BEE2016}.
Therefore to understand the dynamics of real complex systems we often need to describe interactions beyond the simplest pairwise relationships.

Hypergraphs~\cite{B67,B73,J13} naturally encode higher-order interactions in terms of fundamental units, their hyperedges. However, in many cases the data only reveals pairwise interactions making the simple graph the appropriate representation and the higher order interactions have to be inferred from the pairwise interactions recorded in a network.  For instance, data on phone calls does not reveal other higher-order connections, e.g.\ through face-to-face meetings, but the pattern of calls can reveal the existence of such connections.
Network Science offers many approaches to discovering these local group interactions from a network.
It is natural to work with small sub-graphs, motifs \cite{MILO2002,BAD2016},\st {or if we consider induced subgraphs (the maximal subgraph defined by a given vertex set), in terms of} graphlets \cite{PCJ04,P07b}. Higher order analysis in terms of paths is important in several contexts\cite{GL16,S17} but more relevant here is the use of cliques, subgraphs which are complete graphs. Cliques have long played an important role in social science, for instance see \cite{F92}, and can be used in many contexts such as community detection \cite{DPV05,E10}.  Cliques in networks are the basis for analysis in terms of simplicial complexes as used in algebraic topology. In network analysis, simplicial complexes~\cite{MM2006, HDSM2009,CRD2016} have been used to analyse network geometry~\cite{GCE2017}, to model structure in temporal networks~\cite{GAS2018}, investigate synchronization phenomenon~\cite{AJG2018,AJS2019,PAA2019,AJG2020}, social contagion~\cite{IPBL19}, epidemic spreading~\cite{JSA2020}, and neuroscience~\cite{PGP2014,ELKP19}.

%Not sure this is very relevant
%Though the clusters found by community detection methods~\cite{NCOM2012, FD2016} are on much larger scales, the methods are typically based on probing beyond the edge and are often based on higher order interactions in the network \cite{E10}. local generally find larger clusters and are on much larger scales from what we are inter
%Although they have demonstrated the importance of special structures within the complex systems, they cannot explicitly describe group interactions.

% I do not see what bipartite graphs have to do with higher order interactions except in the sense that projections might do this \cite{GJM2006}
% Line graphs \cite{ER2009}

One area where network analysis is less well developed is the temporal evolution of networks. Many complex systems are not static so an important application of networks is to analyse their behaviour over time~\cite{HS12,MAR2013,PRN2013,ENR2014,JRL2015,ML16,OFV2019,OLA2020}. Using higher-order interactions in an evolving network context has been considered in a few contexts~\cite{RIA2013,INH2015,ISW2017,IPBL19}. Most of the research on motifs in temporal networks focuses on how the number of each type of motif changes as the network is evolves. On the other hand, the study of simplicial complexes is interested in how the fully connected triangles affect the structures of networks. So there is a gap between understanding how three-node combinations will evolve and how that evolution will affect the evolution of the whole network. This gap motivates us to design a method to investigate whether any non-pairwise interactions will be observed by measuring three-node dynamics.

In this paper we study higher-order processes in temporal networks through the simplest examples of a higher-order interactions, the evolution of the node triplet, three nodes and all three potential edges between these three nodes. Our method looks at the evolution of a network through the relationships of three nodes \hl{as shown in the top row of}~\Figref{f:triadset}.
% Figure 1.}
\st{, so the higher-order structures include a clique, the triangle, and a path of length two.} Studying the evolution of relationships in a network by focussing on just three nodes has a long history as this includes the process where a path of length two connecting three nodes turns into a triangle as in triadic closure \cite{R53,G73,KH07,GRJ2014} which underpins many other concepts such as structural holes \cite{B04b}.

This paper is organised in the following way: In the first section, we introduce the transition matrix that describes the Markovian evolution of the triplet. In the second section, we propose a null model that can demonstrate the higher-order interactions indeed exist in many real-world networks. In the third part, based on the triplet evolution, we create an algorithm that can predict the existence of the links in the temporal networks.

% ***************************************************************************
\section*{Triplet dynamics}

In order to look beyond pairwise interactions we focus on triplets, \st{at the configurations of three nodes in a network along with all their edges, i.e.  the three-node graphlet (also known as a ``triphlet''). }\hl{ configurations of three nodes in a network along with any edges between those nodes as shown in}~\Figref{f:triadset}.
% Figure 1.
Our focus is not on the number of each triplet type at each time but on how each triplets evolves from one arrangement to the next in a temporal network, an example of ``temporal graphlets''.

%includes what are also called temporal motifs\cite{HS12,ML16} and simplicial complexes\cite{IPBL19}.

% ----------------------------------------------------------------
\subsection*{Transition matrix}

We start with temporal graphs $\Gcal(s)$, a sequence of graphs with one node set $\Vcal$ but with variable edges sets $\Ecal(s)$, where $s$ is a discrete time variable. The variable $\Triad_s(u,v,w)$, often abbreviated to $\Triad_s$, records the state of a\st{ three-node} triplet $(u,v,w)$ at time $s$.  The states are the subgraphs equivalent to the three nodes and all the edges between them at time $s$.  That is $\Triad_s\equiv\Triad_s(u,v,w)$ is a map from a node triplet $(u,v,w)$, where $u,v,w \in \Vcal$, to the induced graphlet, the maximal subgraph in $\Gcal(s)$ containing the nodes $u$, $v$ and $w$. We will use $\Triadset$ to denote the set of all the possible three-node graphs and $m_{i} \in \Triadset$ represents one of the possible graphs in $\Triadset$. So formally
\beq
    \begin{array}{rccl}
       \Triad_{s} \colon & \Vcal\times\Vcal\times\Vcal &\rightarrow & \Triadset ,\\
                         & (u,v,w)                     &\mapsto     & \Triad_{s}(u,v,w) = m_i .
    \end{array}
    %\, .
    \label{e:Tripletdef}
\eeq

\begin{figure}
	\centering
	\includegraphics[scale=0.9]{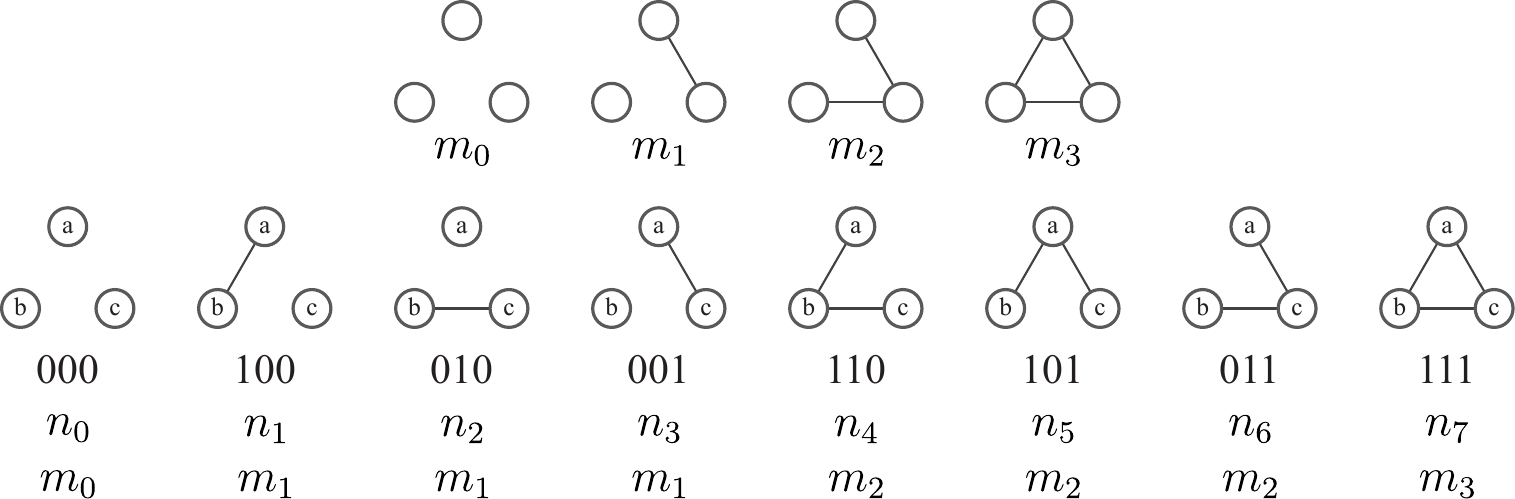}
	\caption[Examples of triplet states in $\Triadset$]{In the top row are the four triplet states of $\Triadset_4$, used only for illustration in this paper. The four states $m_{i} \in \Triadset_4$ ($i = 0,1,2,3$) are characterised by the number of links $i$ in the given three-node graph.
	In the lower row,  there are the eight distinct states of triplets when nodes are distinct (labelled). Below each labelled diagram is the binary representation of the edge set, the label $n_i \in \Triadset_8$ ($i = 0,1\dots7$), and the corresponding $m_i$ triads if labels are ignored.}
		\label{f:triadset}
	\end{figure}

The choice of $\Triadset$ is not unique and it depends on the characteristics of the nodes and links used to distinguish configurations. If we characterise the states by the number of links among the three nodes, that is use unlabelled graphs, there are just four distinct states in $\Triadset$ which we can name as $m_0$, $m_1$, $m_2$ and $m_3$ as shown in ~\figref{f:triadset}. So $\Triadset_4$ is the state set characterised by the number of links and here $m_i$ is the unlabelled graph of three nodes and $i$ edges. We will use $\Triadset_4$ for visualisation and illustration only in our work here.

By contrast, we want to consider the labels (identities) of the nodes in our work. So for our results we work with eight states in $\Triadset$ since each link between the three pairs of nodes can either be present or absent. That means the links between different pairs of nodes are distinct. This state set is represented by $\Triadset_{8}$ as illustrated in the lower row of~\figref{f:triadset}.

Typically one uses graphlets by counting the frequency of each graphlet $m \in \Triadset$ in each graph $\Gcal(s)$ in the temporal graph sequence. So a simple way to look at the evolution is to see how these counts change.  We wish to go beyond this and look at the way local structure controls the local evolution. In order to do this we define a transition matrix $\Tsf$ which describes the likelihood that a given triplet is to be transformed in to another triplet in one time step. That is $\Tsf_{ij}(s)$ gives the probability that a triplet of nodes in the state $m_i$ in $\Gcal(s-1)$ at time $s-1$ becomes the triplet $m_j$ in $\Gcal(s)$ at the next time step, $s$. That is
\begin{equation}
    \Tsf_{ij}(s) =  \Prob (\Triad_{s} = m_j| \Triad_{s-1} = m_i) \, .
    \label{e:Tdef}
\end{equation}
%So the diagonal term $\Tsf_{ii}, i \in \{0,1,2,3\}$ measures the probability of a triplet remaining the same for each transition steps $\Delta t$.
By definition, all entries of $\Tsf$ are non-negative, $\Tsf_{ij} (s) \geq 0$, and each row in the transition matrix satisfies a normalisation condition
\begin{equation}
    \sum_{j} \Tsf_{ij} (s)
    =
    \sum_{m_j \in \Triadset}\Prob (\Triad_{s} = m_j| \Triad_{s-1}=m_i)
    =1 \text{,} ~ \forall \; m_i \in \Triadset \, .
\end{equation}

In practice we have to use an estimate $\Tsfhat(s)$ for the transition matrix $\Tsf(s)$.  We do this by using a subset $\Tripletset_{s-1}$ of all possible distinct node triplets so $\Tripletset_{s-1} \subseteq \Vcal^3$. From this subset of node triplets, we then count how often the associated graphlet transforms from $m_i$ to $m_j$. More formally we define
\begin{subequations}\label{e:Tsfhatdef}
\begin{eqnarray}
 \Tsfhat_{ij} (s)
 &=&
 \frac{1}{k_i}
 \sum_{(u,v,w) \in \Tripletset_{s-1}}
 \delta(\Triad_{s}(u,v,w),m_j) \; \delta(\Triad_{s-1}(u,v,w),m_i)
 \, ,
 \\
 k_i
 &=&
 \sum_j \sum_{(u,v,w) \in \Tripletset_{s-1}}
 \delta(\Triad_{s}(u,v,w),m_j) \; \delta(\Triad_{s-1}(u,v,w),m_i)
\end{eqnarray}
\end{subequations}
where $\delta(\Gcal_1,\Gcal_2)=1$ ($0$) if graphlets $\Gcal_1$ and $\Gcal_2$ are isomorphic (not isomorphic). The best estimate $\Tsfhat_{ij} (s)$ of $\Tsf_{ij}(s)$ is produced if $\Tripletset_{s-1}$ is the set of all possible distinct node triplets. However, for a large graph with $N$ nodes there are $\binom{N}{3} = N\cdot(N-1)\cdot(N-2)/6$ three node combinations making it computationally inefficient to use all triplets. For example, for $N = 100,000, \binom{N}{3} = 1.7\cdot 10^{14}$, Therefore, we will use random sampling of triplets of a sufficient amount to produce our estimates. The estimations are detailed in Appendix A.
% \appref{a:tf_estimation}

\begin{figure}[ht!]
     \centering
     \includegraphics[scale = 0.7]{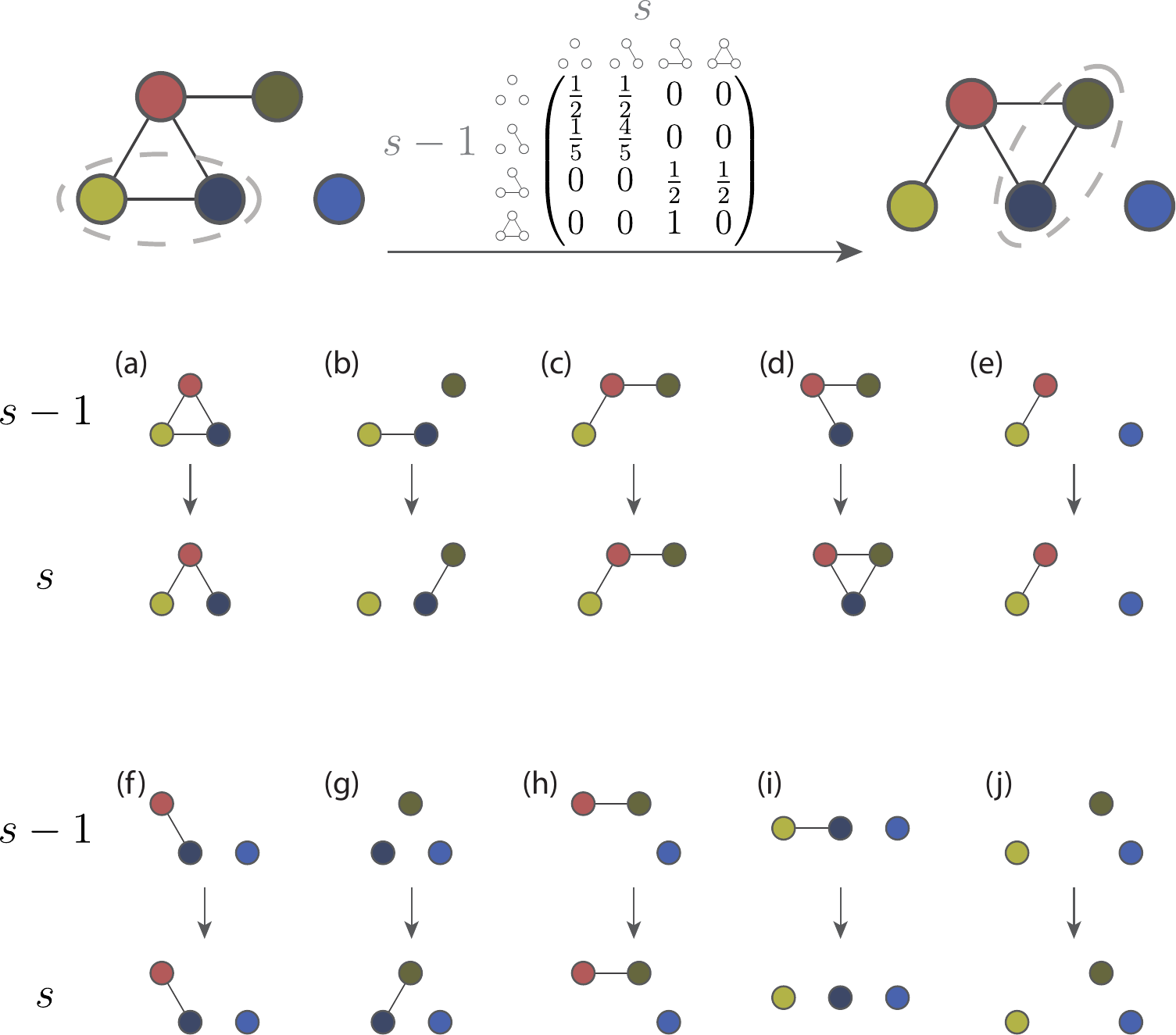}
     \caption{An illustration of how the empirical transition matrix $\Tsfhat(s)$, shown top centre, can be computed from evolution of the network $\Gcal(s-1)$ shown top left, with links at time $s-1$, to $\Gcal(s)$ shown top right, with links at time $s$. When the states are only characterised by the number of links of the three-node combinations, $\Mcal_4$, for illustrative purposes, there are four possible states being denoted by $m_i$ ($i = 0, 1, 2, 3$) of \figref{f:triadset}. The subscript $i$ represents the number of the links $i$ in the graphlet $m_i$, so $m_0$ is the triplet with zero links, $m_{1}$ is a triplet with one link and so on.
     There are $\binom{5}{3} = 10$ ways of choosing $3$ nodes for a set of $5$ nodes, and the evolution of all ten triplets are shown in the remaining rows in the figure above. For instance, the subgraph induced by a triplet (a) is the triplet $m_3$.  This triplet loses an link in the next graphlet, so the same node triplet is now associated with the two link triplet $m_2$.  This change, therefore, contributes to the $\Tsfhat_{32}$ entry. As this is the only induced subgraph (graphlet) in the earlier $\Gcal(s-1)$ graph which is isomorphic to the $m_3$ triad, this means $\Tsfhat_{32}=1$ while $\Tsfhat_{3j}=0$ for $j = 0,1,3$.  On the other hand, the double link $m_2$ triad appears twice as an induced subgraph of node triplets in $\Gcal(s-1)$, namely (c) and (d).  These triplets have two different triads in $\Gcal(s)$ leading to two non zero entries in the $\Tsfhat_{2j}$ row, $\Tsfhat_{22} = \Tsfhat_{23} = 1/2$. Note we use $\Mcal_8$ of \figref{f:triadset} in our analysis.}
     \label{f:trans_mat_ex}
 \end{figure}

To illustrate the construction of a triplet transition matrix, we use the simpler $\Triadset_{4} =  \left \{ m_{0}, m_{1}, m_{2}, m_{3} \right \}$. Note that the subscripts of $\Tsf$, $i,j = 0,1,2,3$ correspond to subscripts of states $m_i$ in $\Triadset_4$ shown in \figref{f:triadset}. An example of the evolution of a network of $5$ nodes and the triplet transition matrix is shown in ~\figref{f:trans_mat_ex}. Note we will use labelled subgraphs and $\Mcal_8$ for our analysis.

% *************************************************************************
\section*{Evidence for higher order interactions}

To investigate the effectiveness of three-node interactions, we start by considering a `pairwise model' whose dynamics is driven only by pair-wise relationships. This will act as a null model when analysing the transition matrix for artificial and real-world networks.

% -------------------------------------------------------------
\subsection*{A pairwise model}~\label{subsec:pairwise_model}
Our `pairwise model' is a stochastic graph model for dynamic networks~\cite{ZX2017} in which the evolution of the links is based on comparison of the evolution of the edge between pairs of nodes, so a two-node graphlet model.  In the pairwise model, moving from one network, $\Gcal(s)$ to the next network in the sequence, $\Gcal(s+1)$, any pair of nodes with no existing link gains a link with probability $p$ otherwise with probability $(1-p)$ the pair of nodes remains unconnected. Similarly, every existing link $e \in~\Ecal(s)$ is removed with probability $q$ otherwise with probability $(1-q)$ the link remains. The number of links is not preserved in this model.
%  though it will tend towards a fraction $q/(p+q)$ of the maximum possible number.
This null model is a lower order description of the local interactions than our full analysis in terms of triplets. It is straightforward to write down the form of the transition matrix in our triplet based analysis when assuming this pairwise model gives a precise description, giving us a two-parameter transition matrix denoted as $\Tsfpw(s)$. The detailed calculation of the form is given in Appendix B.
% ~\appref{a:simple_model}

If we assume that graph evolution follows this pairwise mechanism, we can estimate values $\phat(s)$ and $\qhat(s)$ for the parameters $p$ and $q$ respectively by looking at how links changed over one time step, i.e., from the edge set $\Ecal(s-1)$ in $\Gcal(s-1)$ to edge set $\Ecal(s)$ of $\Gcal(s)$. Formally we have that
\beq
 \qhat(s) = 1- \frac{|\Ecal(s-1) \cap \Ecal(s)|}{|\Ecal(s-1)|} \, ,
 \label{e:qhatdef}
\eeq

\beq
 \phat(s) = \frac{\left|\Ecal(s) \setminus (\Ecal(s-1) \cap \Ecal(s) ) \, \right|}{N(N-1)/2 - |\Ecal(s-1)|} \, .
 \label{e:phatdef}
\eeq
This gives us our pairwise model prediction for the triplet transition matrix $\Tsfhatpw(s)$, where we substitute  $\phat(s)$ from \eqref{e:phatdef} and $\qhat(s)$ from \eqref{e:qhatdef} for $p$ and $q$ in $\Tsfpw$ in equation (B1) of appendix B. That is $\Tsfhatpw(s) = \Tsfpw (\phat(s),\qhat(s)) $ where
\beq
 	\Tsfpw (p,q)
 	\! = \!\!
 	\begin{pmatrix}
 		(1\!\!-\!p)^3 & 3p(1\!\!-\!p)^2 &3p^2(1\!\!-\!p) & p^3 \\
 		q(1\!\!-\!p)^2 & (1\!\!-\!q)(1\!\!-\!p)^2+2qp(1\!\!-\!p) & 2(1\!\!-\!q)p(1\!\!-\!p)+qp^2 & (1\!\!-\!q)p^2 \\
 		q^2(1\!\!-\!p) & 2(1\!\!-\!q)q(1\!-\!p)\!+\!q^2p & (1\!\!-\!\!q)^2(1\!-\!p)+2qp(1\!-\!q) & (1\!-\!q)^2p \\
 		q^3 & 3(1\!-\!q)q^2 & 3q(1\!-\!q)^2 & (1\!-\!q)^3 \\
 	\end{pmatrix}
 \, .
 \label{e:Tsfhatpwdef}
\eeq

% ---------------------------------------------------------------
\subsection*{\st{Quantifying non pairwise interactions}}

\subsection*{\hl{Real World Data Sets}}

The first data set is \st{a} Turkish Shareholder network~\cite{QY}. The nodes are shareholders in Turkish companies and two shareholders are linked if they both hold shares in the same company in one year. The snapshots in the Turkish \st{companies temporal}\hl{Shareholder} network, $\Gcal(s)$, are for consecutive years.
The second data set is a Wikipedia Mathematician\st{s} network. Nodes are pages corresponding to biographies of mathematicians in Wikipedia and two nodes are connected in one snapshot $\Gcal(s)$ if there is a hyperlink (in either direction) between the two pages at the time the data is taken~\cite{BZT19}.  %Here the snapshots are taken at irregular intervals.
The third temporal network is constructed from \hl{c}ollege message data. In this case the nodes are students and an edge in a snapshot indicates that the students exchanged a message within the interval associated with that snapshot~\cite{POC09,OP09,SNAP}.
We also use a network based on face-to-face contacts at a conference on hypertext. In the network, a node represents a conference visitor, and an edge represents a face-to-face contact that was active for at least 20 seconds \cite{Hypertext,KONECT}.
Finally the Email network is derived from the emails at a large institution. Each node corresponds to an email address. An edge in a given snapshot indicates that an email was sent between the nodes in the time interval corresponding to that snapshot~\cite{LKF07,EUEmail,SNAP}.
For temporal networks based on real data, the actual time interval between consecutive graphs in our temporal network, between $\Gcal(s)$ and $\Gcal(s+1)$, is given in real units as $\Delta t$.  In some data sets we are able to look at the same data with different real time intervals between data sets. We provide a summary of the graph statistics in~\Tabref{tab: Summary_of_network}:
\newlength{\timwidth}
\settowidth{\timwidth}{7 days, 1 month}

\begin{table}[htb!]
\centering \footnotesize
\begin{tabular}{|rl||c|c|c|}
\hline
Dataset & (Abbreviation)         & Nodes ($N$)  & Edges in Static Graph ($E$)   &  Temporal Edges ($TE$)   \\  \hline
%%%& & ($N$)  & ($E$)  & ($TE$)     \\ \hline
Wikipedia Mathematician\st{s} & (\texttt{WikiMath})    &   6049 & 15016  &  36315 \\
     Turkish Shareholder & (\texttt{Shareholder}) &   1804 & 20981  &  27218 \\
               Hypertext & (\texttt{Hypertext})   &    113 &  2196  &  20818 \\
         College Message & (\texttt{CollegeMsg})  &   1899 & 20296  &  59835 \\
       Institution Email & (\texttt{Email})       &    986 & 24929  & 332334 \\ \hline
%Hep-Ph  & 28,093 & 4,596,803 & 1993-2003 (yr) & 1 day                \\
\end{tabular}
\caption{Summary of network statistics. Edges in static graph gives the number of node pairs that have at least one interaction while the Temporal Edges gives the total number of interactions between edge pairs, so some edge pairs have more than one temporal edge.
}
\label{tab: Summary_of_network}
\end{table}

\subsection*{Quantifying non pairwise interactions}

The pairwise model captures the effects of the interaction of node pairs. We can use this in the form $\Tsfhatpw(s)$ in \eqref{e:Tsfhatpwdef} as a benchmark to show how higher-order information is captured by our triplet based analysis using $\Tsfhat(s)$ of \eqref{e:Tsfhatdef}.
One way to study this for any given temporal network is to look at the difference $\Delta \Tsfhat(s)$ of the transition probabilities between the empirical triplet transition and the assumed pairwise null model:
\begin{equation}
 \Delta \Tsfhat(s) = \Tsfhat(s) - \Tsfhatpw(s),
 \label{e:deltaTsfhatpwdef}
\end{equation}
where each entry in the matrix represents the difference between the estimated triplet transition probability and pairwise transition probability for different states.

To test our approach using our triplet transition matrices, we first look at artificial temporal graphs created from three simple models.  The first model, the pairwise model, is simply the same pairwise interaction mechanism used above to define $\Tsfhatpw(s)$.  The second model, the edge swap model, is the configuration model which is also based on pairs of nodes. The third model, the random walk model, uses short random walks in order to find new target nodes when rewiring an edge, and so this involves higher-order interactions. In the last two models, a fraction of edges in $\Gcal(s)$ are rewired to give the next graph $\Gcal(s+1)$ in the temporal network. The results are as expected with matrix $\Delta \Tsfhat(s)$ close to zero for all entries for the first two models based on node pairs and it is only with the third higher-order model that some $\Delta \Tsfhat(s)$ entries are found to be large. The networks constructed are undirected. The details of the models and of the results are given in Appendix B.

\begin{figure}
    \centering
    \includegraphics[width=0.95\textwidth]{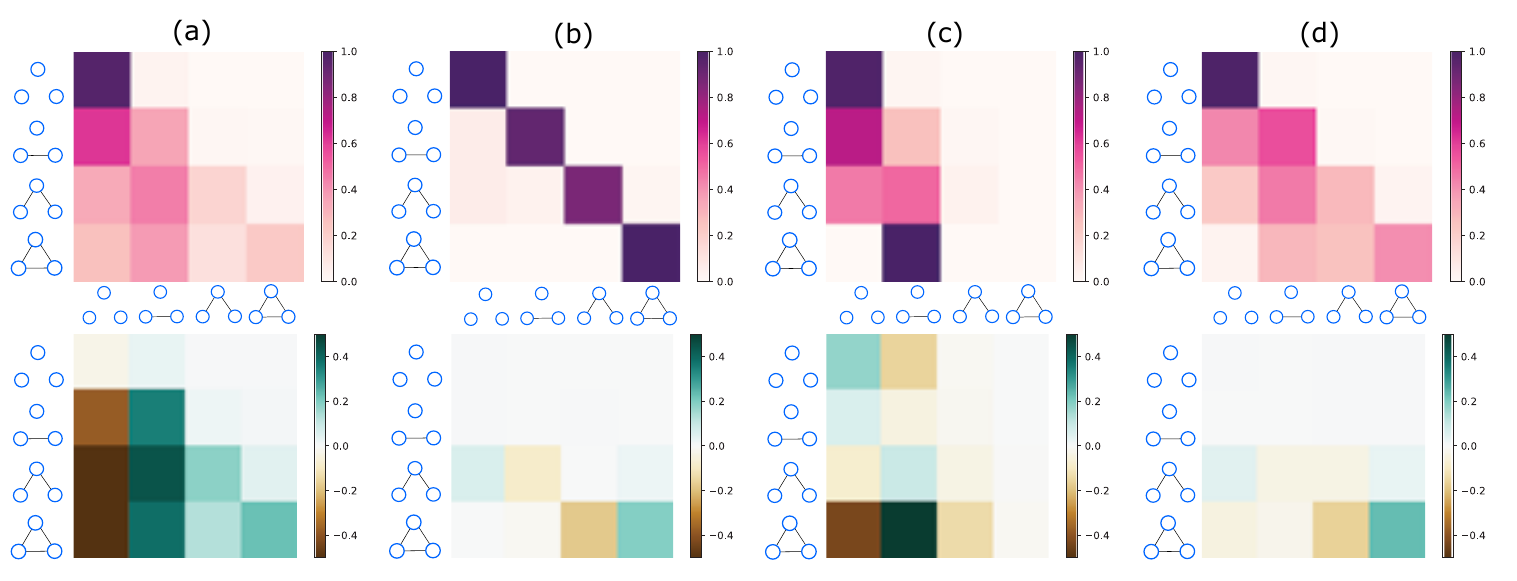}
    % \vspace*{5mm}
    \caption[Triplet transition matrix for real world networks]{The Triplet transition matrix $\Tsfhat$ evaluated for real world networks. The average values $\texpect{\Tsfhat}$ \eqref{e:Tsfhatdef} are shown on the top row while the average of differences from the simple pairwise model, $\texpect{\Delta \Tsfhat}$ of \eqref{e:deltaTsfhatpwdef}, are shown on the bottom row.
    Each of the four-by-four heat map grids is organised in the same way as the $\Tsfhat$ matrix shown in~\figref{f:trans_mat_ex}. That is, the rows indicate the initial triplet state ($m_0$ to $m_3$ from top to bottom as indicated) and the columns indicate the final triplet state ($m_0$ to $m_3$ from left to right as indicated).
    The values of $\texpect{\Tsfhat}$ on the top row run from $0$ (white) to $1.0$ (dark red).  For $\texpect{\Delta \Tsfhat(s)}$ on the bottom row, values run from dark brown (-0.5) through white ($0.0$) to dark blue ($+0.5$).
    Each column of heat maps is for a different data set; from left to right we have: (a) Turkish \hl{S}hareholder network\st{s} ($\Delta t = 2 \mathrm{yr}$), $p=3.49\times10^{-2}$, $q=0.638$, (b) Wikipedia Mathematician\st{s} network\st{s} ($\Delta t = 1 \mathrm{yr}$), \st{$p=5.71\times10^{-4}$}\hl{$p=3.19\times10^{-4}$}, $q=0.068$, (c) College Message network\st{s} ($\Delta t = 1 \mathrm{mo}$), $p=0.067$, $q=0.510$, and (d) Email network\st{s} ($\Delta t = 1 \mathrm{mo}$), $p=9.43\times10^{-3}$,  $q=0.432$. The actual time difference between snapshots $\Gcal(s)$ and $\Gcal(s+1)$ is $\Delta t$.  	
    % ~\appref{a:dataset}. The description of the data sets can be found in Appendix C.
    }
    \label{f:tm_real}
\end{figure}

We then apply this framework to analyse some real data sets.  For clarity we show results for our triplet transition matrix \eqref{e:deltaTsfhatpwdef} when working with $\Triadset_4$ and these are shown in~\Figref{f:tm_real}.
\st{Not surprisingly, given these are from real world data sets, these who significant non-zero entries in our measure $\Delta \Tsfhat(s)$ but also they all look very different from each other, reflecting different mechanisms behind the evolution of different systems.}\hl{These results for real-world systems have significant non-zero entries in our measure $\Delta \Tsfhat(s)$. These results for $\Delta \Tsfhat(s)$ are also very different from each other, reflecting distinct mechanisms behind the evolution of these systems.}

When we look for higher order interactions, we find clear differences between the triplet transition matrix $\Tsfhat$ and the simple pairwise reference model of $\Tsfhatpw$, especially in the Turkish \hl{S}hareholder network \subfigref{f:tm_real}{(a)}. For example, compared to the corresponding probability in the pairwise model, the real probability of any triplet state becoming disconnected in the next snapshot (i.e., moving from $m_i \to m_0$, $i \in \{0,1,2,3\}$, the first column of the transition matrix) is much less, showing that this subgraph is very stable compared with the pairwise case in the \hl{Turkish} \hl{S}hareholder network. Additionally, the state is much more likely to evolve to $m_1$ at $s+1$ (the second column of the transition matrix), which demonstrates that in many real networks, the interactions are beyond the pairwise interaction. The significance test using Z-score can be found in Appendix D.
% ~\appref{a:zscore}.

In our analysis of real data, we use $\Delta t$ to denote the physical time difference between snapshots $\Gcal(s)$ and $\Gcal(s+1)$. This time interval can be varied when working with the College message data of~\subfigref{f:tm_real}{(c)} and the Email network\st{s} of~\subfigref{f:tm_real}{(d)}. In these cases we try several values of $\Delta t$ before choosing an appropriate value the the illustrations in~\subfigref{f:tm_real}. We are looking for a value that is not too short (nothing much happens) and not too long (averaging over uncorrelated changes).

Unsurprisingly, all four real world networks show considerable higher-order effects but there are interesting differences that reveal the processes behind the evolution are likely to have some significant differences.
The $\Tsfhat$ results~\hl{are shown} in the top row \hl{of}~\figref{f:tm_real}. \hl{The result} for the Turkish \hl{S}hareholder network\st{s} in~\subfigref{f:tm_real}{(a)} and the Email network\st{s} in ~\subfigref{f:tm_real}{(d)} look very similar. However, when we compare them to our reference model, the pairwise model, we see large differences, showing that these are very different types of temporal network and using raw values of $\Tsfhat$ can be misleading.

In fact, the networks which are most similar, once simple pairwise processes are discounted, are the Wikipedia Mathematician\st{s} network\st{s}~\subfigref{f:tm_real}{(b)} and the Email network\st{s} ~\subfigref{f:tm_real}{(d)}.  In both cases, processes where an edge is added (upper triangle in the heat map) there is little difference from the pairwise model.  This might be expected for the cases where there is at most one edge in the triplet. Only the $\Delta \Tsfhat_{23}$ entry for the $m_2 \to m_3$ transition shows a slight increase over what would be predicted based on the rate of edge addition in these models (the $p$ parameter of the pairwise model) which suggests triadic closure~\cite{R53,G73,KH07,GRJ2014} does play a role here but it is very slight by these measures. On the other hand, there is a strong sign of ``triadic stability'', that is the complete graphlet $m_3$ is much more stable than we would expect given the rate of edge loss in the pairwise model (the $q$ parameter). \st{So it seems the social processes behind the ideas of triadic closure are possibly more important for the stability of triangles in the contexts of our Wikipedia Mathematician and Email network.}\hl{It is natural to think that processes responsible for triadic closure ($m_2 \to m_3$) would also slow the rate at which such triangles break up ($m_3 \to m_2$) but we do not see this in our result. One interpretation of our results is that the social processes normally invoked for triadic closure}~\cite{R53,G73,KH07,GRJ2014} \hl{can, in some cases, be more important in preventing the breakup of triangles than in the creation of triangles.}

In some ways the similarity between the Wikipedia Mathematician\st{s} network\st{s}~\subfigref{f:tm_real}{(b)} and the Email network\st{s}~\subfigref{f:tm_real}{(d)} is surprising as we might have expected the greatest similarity between the two communication networks, the College \hl{M}essage network\st{s}~\subfigref{f:tm_real}{(c)} and the Email network\st{s}~\subfigref{f:tm_real}{(d)} but that is not what we see in our measures.  In particular, the stability of the complete triangle in the college message is exactly as we would expect based on pairwise measures suggesting different properties in these two communication networks, \st{e.g.}\hl{i.e.} that many of the \hl{c}ollege messages are between actors who do not have strong ties.

However, the main message in~\figref{f:tm_real} is that in almost any data, we find the evolution has clear signals of higher-order interactions playing an important role.

%i.e. $\Delta \Tsfhat_{20}$ and $\Delta \Tsfhat_{21}$ and different behaviours in some entries, for example, $\Delta \Tsfhat_{22}$ and $\Delta \Tsfhat_{23}$, indicating that the evolution of three nodes of a binary tree: on the one hand, it indeed shows that this type of connection is more likely to be closed as a triangle compared to pairwise interactions, on the other hand, expect for transforming into a triangle, (a) is more likely to remain connected at least one link while the (d) is more likely to be disconnected.

%The \subfigref{f:tm_real}{(b)} Wikipedia Mathematicians networks and \subfigref{f:tm_real}{(d)} have similar difference matrices $\Delta \Tsfhat$ yet different transition matrices $\Tsfhat(s)$, suggesting that different dynamics of networks can have similar higher-order interaction patterns.

%
% ***************************************************************************
\section*{Link prediction}

The analysis of our transition matrix of triplets reveals that non-pairwise interaction exists in network evolution. We now ask if these higher-order interaction patterns are essential for network evolution. We investigate this question by performing link predictions for dynamic networks based on the higher-order information stored in our triplet transition matrices.

\subsection*{Triplet Transition Score}

The idea behind our algorithm is that the formation or removal of links between a node pair is encoded in our triplet interactions. The likelihood of a link appearing (or disappearing) between a node pair in a snapshot can be obtained by looking at the triplets containing this node pair and using the triplet transition matrix to see what that suggests about the evolution of any edge between our chosen node pair.

%%%\tcomment{``First, we assume that at each step, one transition occurs for each type of triplet''. TSE:- No I don't think we are assuming this.  Each edge in a snapshot can represent many links added or removed between a node pair in the time period covered by the snapshot.  So the transition matrix covers the case where $0\to 1$ (one edge appears where there was none before) and where $0 \to 1 \to 0 \to 1$ (one edge is added, then removed, then finally added back again). The link we are trying to predict in the next snapshot also represents the same combination of events.  If each snapshot covers the same period of time, then we should be OK.}
%In some data sets, such as communication networks, links exist for a very short time period compared with the interval used to define a single snapshot $\Gcal(s)$ and it is relatively easy for links to be created between a node pair at many different times. This constrains the largest number of transition of the time window for our method.

To keep our analysis simple we will assume that the interval between snapshots $\Delta t$ is about the same size as the appropriate time scale for changes in the network.\st{ If we look at snapshots covering a short period of time, nothing much happens and the transition matrix has too little information in it.  One could then look at larger times scales by using $\Tsfhat(s)$, $\Tsfhat(s-1)$, $\Tsfhat(s-2)$ and so forth to predict links in the next network snapshot $\Gcal(s+1)$. We}\hl{If we look at snapshots covering an extremely short period of time, say one that covers the typical difference in time between events, then the transition matrix contains too little information. In such a case one could then look at larger times scales by using $\Tsfhat(s), \Tsfhat(s-1), \cdots, \Tsfhat(s-n)$ (for some $n$) to predict links in the next network snapshot $\Gcal(s+1)$. However, we} will use the simpler assumption that working with one snapshot by choosing $\Delta t$ appropriately (where we have that choice) captures the essential information.
This implies we have avoided the opposite problem, where $\Delta t$ is too large so the information in the transition matrix is averaging over mostly uncorrelated and unrelated events which means we have very little signal encoded in our transition matrices.

Our second assumption is that the transition depends only on the state of the system at the last time step, which means we assume the process is Markovian. In some ways this need not be completely true.  Provided the snapshots $\Gcal(s-1)$ and $\Gcal(s)$ are in a similar state to the one we are trying to predict, $\Gcal(s+1)$, then because we are constructing our transition matrices based on the similar states, the history of the evolution may well be encoded in this. For instance, patterns of activity in an email network of a large institution could well change if there is a major reorganisation with many people changing roles or locations in which case the history of the system has an impact on the evolution. Put another way, we assume that any non-Markovian behaviour is happening on much larger time scale than we are studying and so we can use a Markovian approximation.

 %More snapshots can be easily included in principle, but doing so will increase the complexity of the method and may decrease effectiveness if the information available is spread too thinly.

Our link prediction algorithm assigns a score to each node pair.  By looking at the distribution of these scores we can separate them into node pairs with low scores, predicting no link in the next snapshot, or a high score meaning this node pair will be connected.  For a given snapshot $\Gcal(s)$, for each node pair, say $(u,v)$, we look at all $(N-2)$ triplets containing that node pair count the number of each triplet falling in each state $m \in \Mcal$.  Normalising this gives us our node-pair state vector $\psi(u,v;s)$ as follows
\beq
 \psi_i(u,v;s)
 =
 \frac{1}{N-2}
 \sum_{w \in \Vcal \setminus \{ u,v\}}
 \delta(\Triad_s(u,v,w),m_i) \, ,
 m_i \in \Mcal \, .
 \label{e:psidef}
\eeq
%which is the probability that if we chose a triplet containing the node pair $(u,v)$ uniformly at random for the set of all such tripolets that we find it in a give state $m$.

Our estimate for the transition matrix $\Tsfhat(s)$, based on $\Gcal(s-1)$ to $\Gcal(s)$ evolution, is to be used to tell us about the evolution of this triplet state distribution $\psi(s)$ in  \eqref{e:psidef}. For instance we estimate that the probability $L_\beta$ that a pair of nodes $u,v$ has an link, $\beta=1$, or no link, $\beta=0$, in the graph $\Gcal(s+1)$ is given by the projection from predicted $\psi(s)$:
\beq
 L_\beta(u,v;s+1)
 =
 \sum_{i,j} \psi_i(u,v;s) \, \Tsfhat_{ij}(s) \,  \Pout_{j\beta},
 \label{e:Ldef}
\eeq
where $\Pout_{j\beta}$ is a projection vector for the triplet evolution and $\sum_{\beta=0}^{1} \Pout_{j\beta}=1$. With the other definition $\sum_{j} \Tsfhat_{ij} =1$, and $\sum_{i} \psi_i(u,v;s)=1$, then we have that  $L_\beta$ is properly normalised $\sum_{\beta=0}^{1} L_\beta(u,v) =1$. It is this $L_\beta$ in~\eqref{e:Ldef} that we use as a score for link prediction.

Take the case of $\Triadset_4$ as an example, the projection from triplet distribution onto links uses
\beq
 \Poutmat =
 \begin{pmatrix}
 1   & 0 \\
 2/3 & 1/3 \\
 1/3 & 2/3 \\
 0   & 1 \\
 \end{pmatrix}.
 \label{e:Poutmatdef}
\eeq
The factors here arise for the state set $\Triadset_4$ since the unlabelled graphlets do not distinguish which links are occupied in the one-link triplet $m_1$ or two-link triplet $m_2$.
For the state set $\Triadset_8$ which we use in most of our work, this projection matrix is much simpler with entries either $0$ or $1$. If we choose $(u,v)$ to be the first link, so it is the only link in the triplet we call $n_1$ in~\figref{f:triadset}, then we can use bitwise logical operators to represent $\Poutmat_{j\beta}$ as $ (j \; \mathrm{AND} \; 1)\;  \mathrm{XOR} \; \beta$.

\subsection*{Node Similarity}\label{a:nodesimilarity}

We are using link prediction as a way to test that our triplet transition matrices $\Tsfhat$ capture important higher-order interactions in the evolution of temporal networks. In order to see how effective our approach is, we need to compare against other methods of link prediction. All the methods we use are  listed in~\tabref{t:linkprediction}. All these methods can be discussed in terms of a node similarity score and in this section we will start our examination of these methods by considering the node similarity scores used in each method. This will allow us to examine the relationships between these various methods and ask if they capture higher-order interactions to any extent. The next stage is to turn these similarity scores into a link prediction and we will look at this step in the next section.

\begin{table}[htb!]
\begin{center}
%%% This version for Sci Rep when citations are superscripts
\begin{tabular}{r l||  c| c}
Abbreviation & \textit{Method}${}^{\text{Reference}}$        & Length Scale & \texttt{Code}\\ \hline\hline
AAI  & {\textit{Adar-Academic Index}\cite{LK07} }         & $2$          & \texttt{nx}  \\
CN   & {\textit{Common Neighbour}\cite{LK07} }            & $2$          & \texttt{nx}  \\
EE   & {\textit{Edge Existence}   }                       & $1$          & \texttt{-}   \\
JC   & {\textit{Jaccard Coefficient}\cite{LK07} }         & $2$          & \texttt{nx}  \\
Katz & {\textit{Katz}\cite{LJZ09a} }                      & $\infty$     & \texttt{own} \\
LLHN & {\textit{Local Leicht-Holme-Newman}\cite{LHN06} }  & $2$          & \texttt{own} \\
LPI  & {\textit{Local Path Index}\cite{ZLZ09}  }          & $3$          & \texttt{own} \\
MFI  & {\textit{Matrix Forest Index}\cite{CS97} }         & $\infty$     & \texttt{own} \\
PA   & {\textit{Preferential Attachment}\cite{LK07} }     & $2$          & \texttt{nx}  \\
RAI  & {\textit{Resource Allocating Index}\cite{ZLZ09} }  & $2$          & \texttt{nx}  \\
TT   & {\textit{Triplet Transition}}                      & $\infty$     & \texttt{own} \\
\end{tabular}
\end{center}
\caption{Table of the link prediction methods used and their abbreviations. The length scale given indicates the longest path length involved in the method or equivalently the largest power of the adjacent matrix involved in the method. Under \texttt{code},  \texttt{nx} indicates that a \texttt{NetworkX} \cite{networkx} library routine was used, while \texttt{own} indicates the authors' own code was used. The Edge Existence (EE) approach was not investigated numerically but was included for the sake of comparison. All these methods, except for the Triplet Transition method, have a temporal path length of $0$.  That is they are derived solely from the current snapshot, $\Gcal(s)$, when making a prediction for links in the next snapshot $\Gcal(s+1)$. The Triplet Transition method has a temporal path length of $1$ as it uses $\Gcal(s)$ and $\Gcal(s-1)$ to predict $\Gcal(s+1)$.}
\label{t:linkprediction}
\end{table}

The various link prediction methods can be categorised based on the type of information used to make a prediction about each node pair: those which use only local interactions probing a fixed distance from the node pair, and those global methods which use nodes arbitrarily far from the node pair of interest. This is indicated by the length scale of methods indicated in \tabref{t:linkprediction} and will be clear from the power of the adjacency matrix $\Amat$ appearing in the similarity scores defined in this section.

\begin{figure}[htb!]
    \centering
    \includegraphics[width=0.5\textwidth]{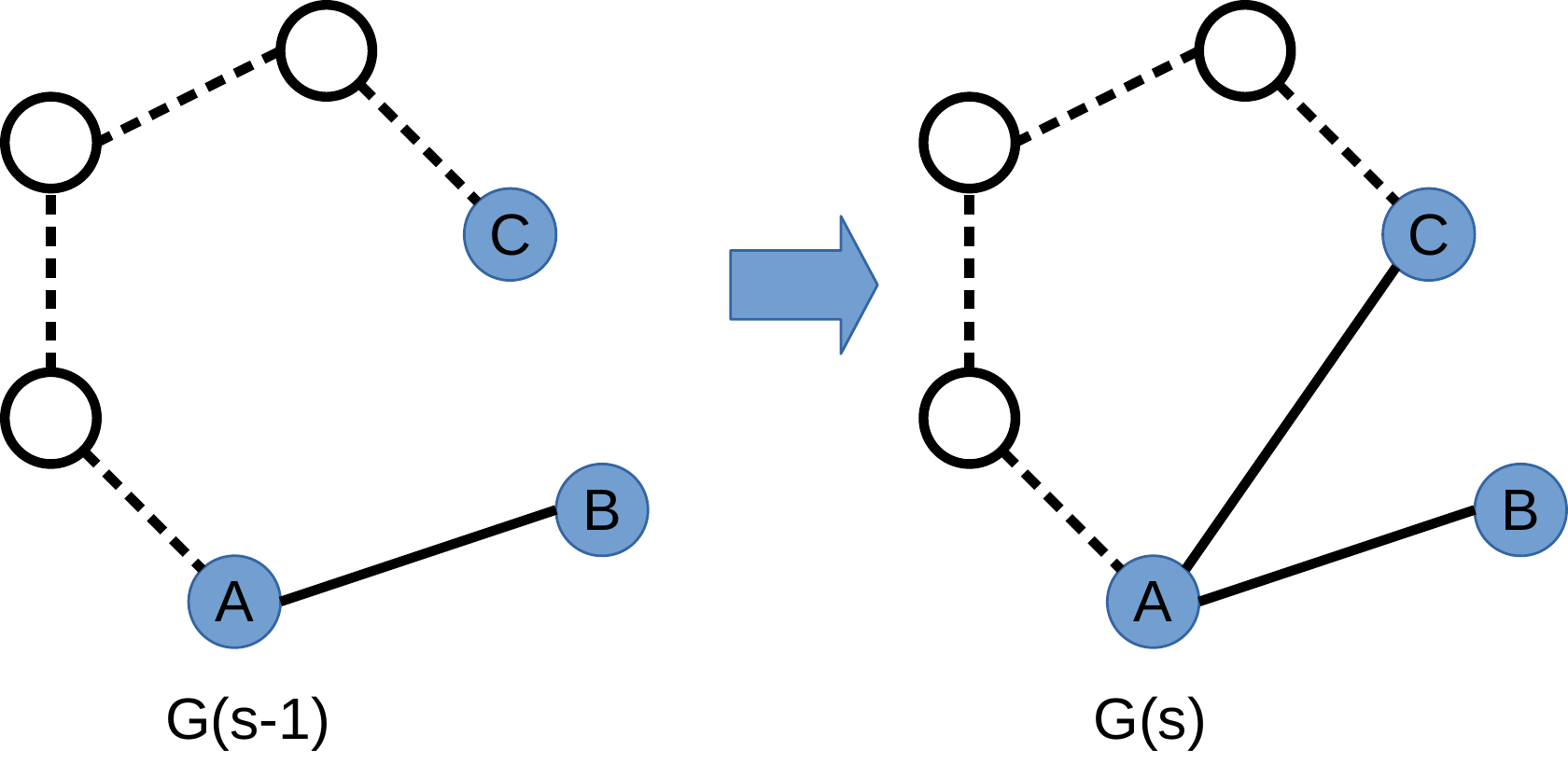}
    % \vspace*{5mm}
       \caption[Long distance paths in the Triplet Transition method.]{An illustration of how information on long range paths between nodes plays a role in our Triplet Transition method. This shows the evolution of one $m_1$ triplet in snapshot $G(s-1)$ to become an $m_2$ triplet in the next snapshot. The triplet nodes are shown as solid blue circles connected by solid black edges. The network outside the triplet is shown with nodes as open circles connected by edges shown as dashed lines. It is the network outside the triplet which provides a long distance connection between triplet nodes $A$ and $C$. Within the original $ABC$ triplet in snapshot $G(s-1)$, $C$ appears disconnected. By using two snapshots, our transition matrix includes the effects of such long-range links and so this approach can explain why the edge $(A,C)$ might be appear more often than otherwise expected from the triplet subgraph alone.}
   \label{f:ttlongrange}
\end{figure}

One other point can be made.  In terms of the temporal network, none of these existing similarity scores, none of the these link prediction methods from the literature, use more than one temporal snapshot $\Gcal(s)$. So one standout feature of our method is the use of two temporal snapshots, $\Gcal(s)$ and $\Gcal(s-1)$. The evolution of a triplet from one snap shot to another records, in an indirect way, the effects of other nodes beyond the triplet of interest as illustrated in \figref{f:ttlongrange}. This is why our method is listed as probing long length scales in \tabref{t:linkprediction}. \hl{So one standout feature of our method is the use of two temporal snapshots, $\Gcal(s)$ and $\Gcal(s-1)$. The evolution of a triplet from one snap shot to another records, in an indirect way, the effects of other nodes beyond the triplet of interest. This is why our method is listed as probing long length scales in }\tabref{t:linkprediction}.

In the following $s(u,v)$ is a (similarity) score assigned to a pair of vertices $u,v \in\Vcal$ where $\Ncal(u)$ is the set of neighbours of vertex $u$, i.e.\  $\Ncal(u) = \{ w | (u,w) \in \Ecal\}$.  The similarity scores are then used to decide if an edge should exist between $u$ and $v$ and that will be used in turn to make the link prediction for the next snapshot in time. We will assume a simple graph in these discussions.

The simplest node similarity measure is the \tdef{Edge Existence} (EE) index $s_\mathrm{EE} (u,v)$.  After all, if there is already an edge between two vertices $u$ and $v$ this is a good indication of a close relationship between these nodes, and vice versa. We define Edge Existence index to be
\beq
 s_\mathrm{EE} (u,v) = A_{uv} = \sum_{e \in \Ecal} \delta_{e,(u,v)} \, .
 \label{e:EEmethod}
\eeq
\st{That is} \hl{E}very edge between $u$ and $v$ adds one to this index so, for a simple graph, this is the corresponding entry of the adjacency matrix, $A_{uv}$. This is the simplest edge prediction method in that it predicts no changes at all so it is not very useful and we do not use it in our work.  However, we will see this score contributes to some of the other, more sophisticated methods, so it is useful to define this score. The Edge Existence score records no higher-order effects, there is a path of length at most one between the nodes.

While not very useful, this \tdef{Edge Existence} (EE) index could produce some very good statistics.  For instance, if very few edges are changing in each time interval, then this method would predict the behaviour of the vast majority of edges as most remain unchanged.  It would only do badly if we used statistics that specifically measured the success of node pairs changing their connectedness from one time step to the next. A good example of\st{this}\hl{ when the Edge Existence method would appear to be successful is  when we break up the data into small steps in time where few edges change. However we would learn little of interest in this case.}\st{ would be when we are able to break the data up into arbitrarily sized steps in time, and if we choose steps which are too small, we will get little change step because of this discretisation choice.}

The \tdef{Common Neighbours} (CN) method~\cite{LK07,LJZ09a,AMA20} simply scores the relationship between two nodes based on the the number of neighbours they have in common
\beq
 s_\mathrm{CN}(u,v) = \sum_{w\in \Vcal} A_{uw} A_{wv} = | \Ncal(u) \cap \Ncal(v) | \, .
\label{e:CNmethod}
\eeq
This will tend to give large scores if $u$ and/or $v$ have high degrees and this is illustrated in more detail in Appendix E.

One way to compensate for the expected dependence of $s_\mathrm{CN}$ on the degree of the nodes is to normalise by the total number of unique neighbours.  This gives us the \tdef{Jaccard Coefficient}~\cite{LK07,AMA20}  (JC) method based on the well known similarity measure~\cite{SM83}
in which the likelihood that two nodes are linked is equal to the number of neighbours they have in common relative to the total number of unique neighbours.
\beq
 s_\mathrm{JC}(u,v)
 = \frac{|\Ncal(u) \cap \Ncal(v)|}{|\Ncal(u) \cup \Ncal(v)|}
 \label{e:JCmethod}
 =
 \frac{s_\mathrm{CN}(u,v)}{k_u+k_v -2s_\mathrm{EE}(u,v) - s_\mathrm{CN}(u,v)}
  \, .
 %\label{e:JCmethod2}
\eeq

The \tdef{Preferential Attachment} (PA) method for link prediction \cite{LK07,AMA20} is based on the idea that the probability of
a link between two vertices is related to the product of their degrees of the two vertices
\beq
 s_\mathrm{PA}(u,v) = |\Ncal(u)| \, \cdot \,  |\Ncal(v)| \, .
 \label{e:PAmethod}
\eeq
This is proportional to the number of common neighbours expected in the Configuration model \cite{MR95} as has been used in the context of collaboration (bipartite) collaboration graphs \cite{N01a,BJNRSV02,LK07}.

The \tdef{Resource Allocating Index}~\cite{ZLZ09} (RAI) method and the \tdef{Adamic-Adar Index}~\cite{LK07,AMA20} (AAI) method are both based on the idea that if two vertices $u$ and $v$ share some `features' $f$ that is very common in the whole network then that common feature is \emph{not} a strong indicator that the two vertices should be linked. The converse is true if the common feature is rare, that is a good indicator that the vertices $u$ and $v$ should be linked. So in general if $W(x)$ is a monotonically decreasing function of $x$ we can use this on the frequency $n(f)$ of the occurrence of feature $f$ to give a generic similarity function of the form $s_\mathrm{W}(u,v) =\sum_{f \in u, v} W(n(f))$. In our case, we will not assume any meta-data exists, but we will look for methods that use features which are based purely upon the topology of the network.

%The \tdef{Resource Allocating index} (RAI) method and the \tdef{Adamic-Adar Index} (AAI) method differ in their choice of weighting function $W$ in defining their similarity score.
In the Resource Allocating Index~\cite{ZLZ09} the inverse of the degree of the neighbour is used as the weighting function, so $W(n(f)) \equiv 1/|\Ncal(w)|$ giving
\beq
 s_\mathrm{RAI}(u,v) =\sum_{w \in {\Ncal(u) \cap \Ncal(v)} } \frac{1}{|\Ncal(w)|} \, .
 \label{e:RAImethod}
\eeq
Note that this means that the contribution from any one node $w$ to the total of all scores $S_\mathrm{RAI}(u,v) = \sum_{u,v} s_\mathrm{RAI}(u,v)$ is half of the degree of $w$ minus one, $(\Ncal(w)-1)/2$. Thus the {Resource Allocating Index} still gives high degree nodes more weight.

On the other hand, the Adamic-Adar Index~\cite{LK07,AMA20}  uses the inverse logarithm of the degree to weight the contribution of each common neighbour to the score,
$W(n(f)) \equiv 1/\ln(|\Ncal(w)|)$. That is
\beq
 s_\mathrm{AAI}(u,v) = \sum_{w \in {\Ncal(u) \cap \Ncal(v)} } \frac{1}{\ln( |\Ncal(w)| ) } \, .
\label{e:AAmethod}
\eeq

All the indices mentioned above have been based either on the degree of the two nodes of interest $s_{EE}$ or on the properties of $u$, $v$ and their common nearest neighbours $w \in \Ncal(u) \cap \Ncal(v)$. This involves paths between the two nodes $u$ and $v$ of length two or less. The next logical step is to include paths of length three and the \tdef{Local Path Index} (LPI) $s_\mathrm{LPI}$ \cite{ZLZ09,LJZ09a} is an example of this where
\beq
 s_\mathrm{LPI}(u,v) = [\Amat^2 + \beta \Amat^3]_{uv}= s_\mathrm{CN}(u,v) + \beta [\Amat^3]_{uv}
  \, .
\label{e:LPImethod}
\eeq
Here $\beta$ is a real parameter where $\beta=0$ reproduces the Common Neighbours score of \eqref{e:CNmethod}. The $\beta\Amat^3$ term is counting the number of walks of length three that start at $u$ and end at $v$. If there is already an edge between $u$ and $v$ then  $[\Amat^3]_{uv}$ includes backtracking paths such as the sequence $u,w,u,v$. This means the second term also includes a term equal to $A_{uv}(|\Ncal(u)|+|\Ncal(v)|)$, i.e.\ there is a contribution from the Edge Existence similarity $s_\mathrm{EE} (u,v)$ \eqref{e:EEmethod} in this method.

The \tdef{Katz Index}~\cite{LK07,LJZ09a,AMA20} (Katz) counts the number of paths between each pair of vertices, where each path of length $\ell$ contributes a factor of $\beta^\ell$ to the score.  The score is simply the appropriate entry of the matrix $[\unitmat-\beta \Amat]^{-1}$,
\beq
 s_\mathrm{Katz}(u,v) = ([\unitmat-\beta \Amat]^{-1})_{uv} \, ,
\label{e:Katzmethod}
\eeq
where $\beta$ is positive but  must be less than the largest eigenvalue of the adjacency $\Amat$. Note for low $\beta$ and for a simple graph we have that %for $u \neq v$ that}
\beq
 s_\mathrm{Katz}(u,v)
 =
 \beta A_{uv} + \beta^2 \sum_w A_{uw}A_{wv} + \beta^3 \sum_{w,x} A_{uw}A_{wx}A_{xv} +  O(\beta^4)
 \label{e:Katzmethodlowbeta1}
 =       \beta s_\mathrm{EE}(u,v) + \beta^2 s_\mathrm{LPI}(u,v) + O(\beta^4)
        \quad \mbox{for } u \neq v\ .
 %\label{e:Katzmethodlowbeta2}
\eeq

The \tdef{Local Leicht-Holme-Newman Index}~\cite{LZ11} (LLHN) is based on the vertex similarity index of~\cite{LHN06}, but while this gives a specific motivation for the form, it is in the end just a specific rescaling of the Katz index \eqref{e:Katzmethod}, namely
\beq
 s_\mathrm{LLHN}(u,v)
 =
 \frac{s_\mathrm{Katz}(u,v)}{s_\mathrm{PA}(u,v)}
 \label{e:LLHNmethod}
 =
 \frac{s_\mathrm{Katz}(u,v)}{|\Ncal(u)| \, |\Ncal(v)|}
 %\label{e:LLHNmethod1}
 = \Dmat^{-1} \left(\unitmat - \beta \Amat\right)^{-1} \Dmat^{-1}
 %\label{e:LLHNmethod2}
\eeq
where $\Dmat$ is a diagonal matrix whose entries are equal to the degrees of the nodes, $D_{uv} = \delta_{u,v} |\Ncal(u)|$.
The motivation for using this normalisation is that $|\Ncal(u)| \cdot |\Ncal(v)|$ is proportional to the number of neighbours expected in the configuration model, as shown in Appendix E.
%%%as shown in~\eqref{ae:ncnrndg}.
So the Local Leicht-Holme-Newman Index is the Katz score relative to the Katz score expected for the same pair of nodes in the configuration model.

The \tdef{Matrix Forest Index}~\cite{CS97} (MFI)  is defined as:
\beq
 s_\mathrm{MFI}(u,v) = [( \unitmat + \Lmat )^{-1}]_{uv} \, ,
 \label{e:MFImethod}
\eeq
where $\Lmat = \Dmat - \Amat$ is the Laplacian.
One way to understand the Matrix Forest Index is to consider the diffusion process described by a Laplacian.  If we were to demand that at time $t+1$ we had only had particles at one site $u$, then $s_\mathrm{MFI}(u,v)$ tells us how many of those particles were at vertex $v$ at the previous time step.

% --------------------------------------------------------------------
\subsection*{From node similarity to link prediction}\label{ss:method}

All the link prediction methods used in this paper, see \tabref{t:linkprediction}, assign a similarity score between pairs of nodes. To turn this into a link prediction, the basic conjecture is that the higher the similarity between a pair of nodes, the more likely we are to find a link between these two nodes.

All methods, therefore, require a precise method to turn the scores into predictions, essentially to define what is meant by a `high' or a `low' score by introducing a \tdef{classification threshold}. Often this is done very simply by ranking the scores and using a fixed number of the most highly ranked node pairs to predict a link. This method is typically used only for \tdef{link addition} in which one is only trying to predict when an unlinked node pair gains a link, that is the $A_{uv}(s)=0$ to $A_{uv}(s+1)=1$ process.

We are interested in the most general predictions, looking at all four possible changes for node pairs from one snapshot in time  to the next, that is all the four possible $A_{uv}(s)=0,1$ to $A_{uv}(s+1)=0,1$ processes. This is \tdef{link evolution} rather than link addition. In order to make these more general predictions for any method, we use $k$-means clustering methods to separate the prediction scores produced by each method into two classes: a high score group and a low score group of node pairs. Any node pair with a score in the high scoring group will be predicted to have a link in the next snapshot; node pairs in the low scoring group will be predicted to have no link.

\begin{figure}[htb!]
    \centering
    \includegraphics[width=\textwidth]{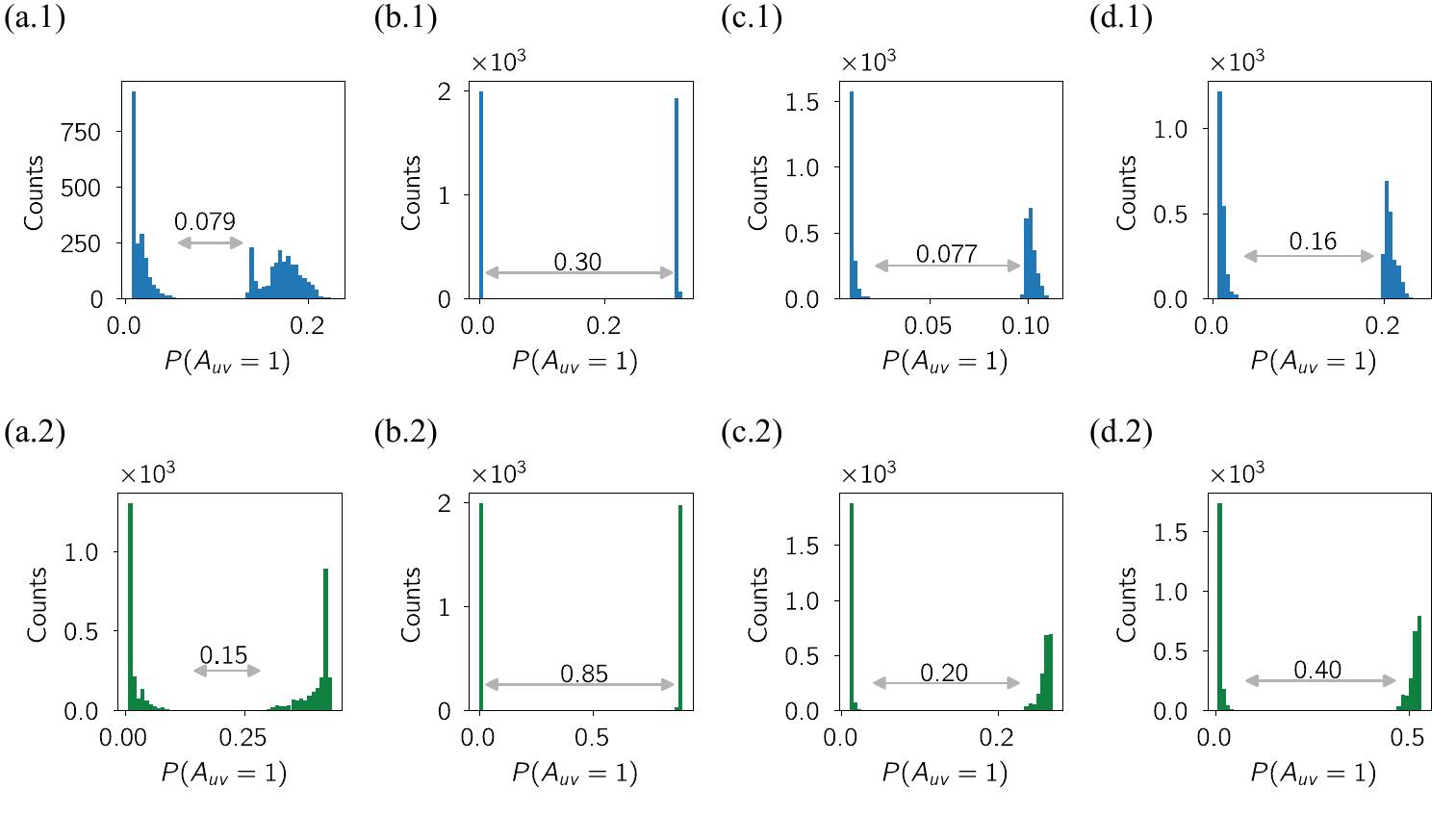}
    % \vspace*{5mm}
   \caption[The histogram of node similarity scores in TT for real data]{The histogram of node similarity scores in our triplet transition (TT) method for (a) Turkish Shareholder network\st{s}, (b) Wikipedia Mathematician network\st{s}, (c) College Message network\st{s}, and (d) Email network\st{s}.
   The precise values are not important as the important feature is the clear separation of the node similarity scores into two groups. One cluster $c_0$ appears to have a `low' probability that a link will exist in the next snapshot and the other cluster has `high' probability. We therefore predict that if the node pair has a score in the lower score cluster $c_0$, then a link will not exist in the next snapshot. Conversely, if a node pair exists in the higher score cluster $c_1$, then we predict a link will exist between this node pair in the next snapshot. The first row, where the histogram is plotted in blue is the clustering results for the $\Triadset_4$ while the second row, where the histogram is plotted in green is the clustering results for the $\Triadset_8$. It shows that the $\Triadset_8$ has higher separation between two clusters than $\Triadset_4$. The results for each network are based on samples which contain at least $2000$ node-pairs with an initial link and at least $2000$ more node pairs which started without a link.}
   \label{f:link_pred}
\end{figure}

To show how this works, we give some examples of how the score from our triplet transition method produces a natural split into low and high scores which is easily discovered by an automated clustering method such as $k$-means, as seen in the first-row of~\figref{f:link_pred}.
In our final results below for our link prediction algorithm, we use $\Triadset_8$  and the clear separation of node similarity scores into a low and high group is shown in the second row of~\Figref{f:link_pred}.
% \hl{Figure 5}.
For comparison we also show the results for our method using the less sensitive $\Triadset_4$. In practice, both seem to separate low and high scoring node pairs well but we get a slightly clearer separation in some cases when using $\Triadset_8$.

\begin{figure}[htb!]
    \centering
    % Put this figure back in
    \includegraphics[width=\textwidth]{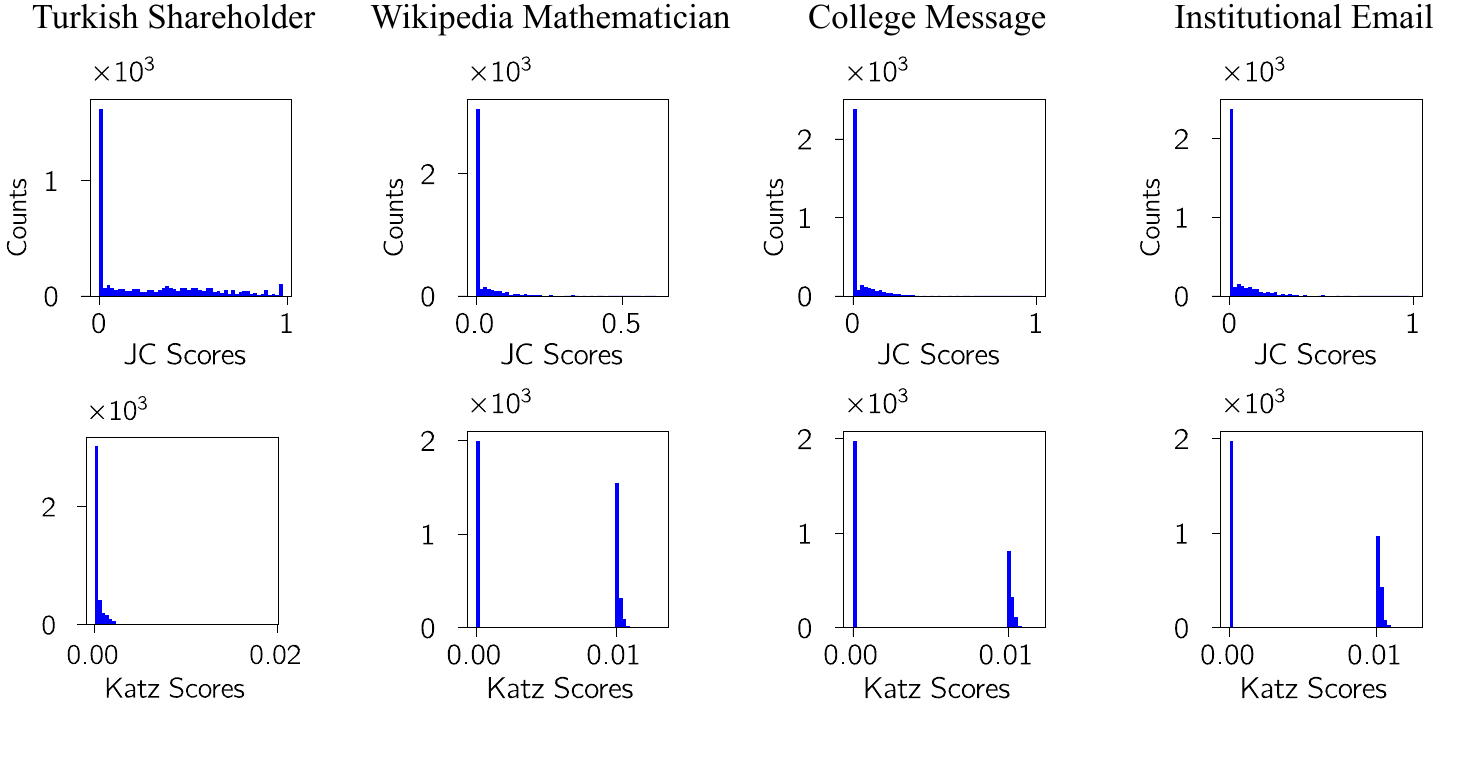}
    % \texttt{otherhist}
    \caption[The histogram of node similarity scores in JC \& Katz methods for real data]{The histogram of node similarity scores using Jaccard Coefficient (JC) \& Katz methods on four real data sets.  Each column is for one data set which are, from left to right: Turkish \hl{S}hareholder network\st{s}, \st{Wiki}\hl{Wikipedia} Mathematician network\st{s}, College Message\st{s} network, and Email network\st{s}. The first row is the similarity score used in Jaccard Coefficient method and the second is for the Katz method. The precise values are not important here as the key feature is the success or failure to identify two clear groups of low- and high-scoring node pairs.
    We under-sampled $1000$ linked node pairs and $1000$ non-connected node pairs\cite{BR03}. Only the Katz method (bottom row) shows the necessary clustering of scores needed for link prediction. Similar plots for all the link prediction methods used in this paper are shown in Figure~E11 of the Appendix.}
    \label{f:other_hist}
\end{figure}

We can look at how the other nine link prediction methods perform in terms of identifying two clear groups of low- and high- scoring node pairs. What is surprising is that most of the algorithms fail to do this well, or some cases at all, in most of the data sets as shown in Section E of the Appendix. An example of a poor separation, the Jaccard Coefficient method, is shown in~\figref{f:other_hist}. The one exception is the Katz method, also shown in \figref{f:other_hist}, which works well in three of the four data sets. Given this is one of the methods probing large distances in the network, this would seem to support the idea that higher-order interactions are important in link prediction.
% ~\figref{f:other_hist} in~\appref{a:othermethods}.

Since our triplet transition method splits the node similarity scores so well, any unsupervised clustering method should be able to split the node pairs into a low scoring cluster and a high scoring cluster. As the problem is in one-dimension a version of k-means clustering is sufficient and this will always assign our node pairs to either a low or to a high score even if a method does not have a clear threshold visually. The objective function of our clustering is $J$ where
\begin{equation}
	J (s_\mathrm{th})
	=
	\sum_{(u,v)}
	\Big[
	        \theta\big(s_\mathrm{th} - s(u,v) \big) \cdot \big|s(u,v) -\mu_- \big|
	\, + \, \theta\big(s(u,v)- s_\mathrm{th}  \big) \cdot \big|s(u,v) -\mu_+ \big|
	\, \Big]
    \, ,
	\label{e:kmeans}
\end{equation}
 \hl{where $\theta$ is the Heaviside step function.} The sum over $(u,v)$ is over all distinct nodes pairs (so $(u,v) \in \Vcal^2 | u \neq v$). The $\mu_+$ ($\mu_-$) is the average similarity score of node pairs in the high score (low score) cluster. The problem is reduced to finding the threshold value $s_\mathrm{th}$ that minimises $J$.

%%%\tcomment{Actually, as this is in one-dimension, the scores are simply real numbers, it is surely quick to sort the node pairs in the order of their scores, then a binary search will find a definitive $s_\mathrm{th}$ very quickly in $O(\ln(N))$? All the pairs  in the low (high) score cluster will be below (above) $s_\mathrm{th}$.}

% ---------------------------------------------------------------------------
\subsection*{Evaluation metrics}

To evaluate the performance of our Triplet Transition and the other link prediction methods, we use a variety of standard metrics.\st{The simplest metric we use is to simply look at the fraction of edges that are removed}\hl{The simplest metric we use is the fraction of edges that are removed} from snapshot $s$ to the next snapshot $s+1$. We denote this by $f_\Ecal (s)$,
\begin{equation}
	f_\Ecal (s)
	=
	1 - \frac{|\mathcal{E}(s+1) \cap \mathcal{E}(s)|}{|\mathcal{E}(s)|} \, ,
	\label{e:flinkchange}
\end{equation}
where $\Ecal(s)$ is the link set of the network in snapshot $s$.

We also use more traditional metrics such as precision and area under the curve for which we need a binary classification. Thinking of snapshot $\Gcal(s+1)$ as the current state while we are trying to predict the state in the next snapshot, we consider four transitions $(A_{uv}(s) \ra A_{uv}(s+1) )$ of node pairs: $0 \ra 0$, $0 \ra 1$, $1 \ra 0$, and $1 \ra 1$. We map these states onto a binary set of states, namely $A_{uv}(s+1)$, so what we call a positive result (negative result) is where a link is present (is not present) in snapshot $(s+1)$ regardless of the state that pair started in. We then consider whether the prediction for the state of the node pair in snapshot $s+1$ was true or false. So a false positive is where we predict no link for a node pair when that pair did end up with a link, while a true negative is where we correctly predict a pair of nodes will not be connected in the next snapshot. The table of predicted classes and actual classes is shown in~\tabref{tab:confusion_matrix}.
\begin{table}[htb!]
    \centering
    \begin{tabular}{|cc|c|c|}
    \hline                                                        &     & \multicolumn{2}{c|}{Actual Classes} \\ \cline{3-4}
         &     & $A_{uv}(s+1) = 1$             & $A_{uv}(s+1) = 0 $             \\ \hline
    \multicolumn{1}{|c|}{\multirow{2}{*}{Predicted Classes}} & $A_{uv}(s+1) = 1$ & True positive ( $01+, 11+$)               & False positive ( $01-, 11-$)               \\ \cline{2-4}
    \multicolumn{1}{|c|}{}                                   & $A_{uv}(s+1) = 0 $ & False negative ( $00-, 10-$)               & True negative ( $00+,10+$)\\
    \hline

    \end{tabular}
    \caption[Confusion matrix for link prediction.]{The Confusion matrix for the prediction of edges between nodes $u$ and $v$ in the next snapshot, $(s+1)$. The adjacency matrix at snapshot $s$ in the temporal network is $\Amat(s)$.  Each of the four outcomes comes from two situations as this confusion matrix does not consider the state of the edge in snapshot $\Gcal(s)$. The two states are shown for each of outcome in brackets with the notation that $\alpha\beta\pm$ shows the value of $A_{uv}(s)$ as $\alpha$, the predicted  $A_{uv}(s+1)$ as $\beta$, and a $+$ symbol (a $-$ symbol) indicates that the prediction made was correct (incorrect).}
    \label{tab:confusion_matrix}
\end{table}

With this traditional binary classification, we can then evaluate the performance of our Triplet Transition method using standard metrics (see Appendix F
% \appref{as:measures}
for more details) --- area under curve and precision. To express these, we define $N_{\alpha\beta\pm}$ to be the number of node pairs satisfying the following criteria. The node pair starts in snapshot $\Gcal(s)$ in state $\alpha$ equal to $1$ if the node pair is connected by an edge, $0$ otherwise. This edge pair is then in state $\beta$ in snapshot $\Gcal(s+1)$ with $\beta$ is $1$ if the node pair is connected by an edge in snapshot $\Gcal(s+1)$, and is $0$ otherwise. Finally the sign indicates if the prediction made for that node pair in $\Gcal(s+1)$ was correct (true, $+$) or incorrect (false, $-$).

We can evaluate the methods independent of the classification threshold chosen through k-means by using the \tdef{area under the curve} (AUC) where the curve is the receiver operating characteristic curve. For any binary classifier of the results of an algorithm, such as defined here in \tabref{tab:confusion_matrix}, the curve plotted is the fraction of positive results (TPR, true positive rate) which are correct against the fraction of negative results which are incorrect (FPR, false positive rate) as the threshold $s_\mathrm{th}$ is varied:
\begin{equation}
	\mathsf{TPR}
	=
	\frac{N_{11+}+N_{01+}}{N_{11+}+N_{01+}+N_{00-}+N_{10-}} \, ,
	\quad
	\mathsf{FPR}
	=
	\frac{N_{01-}+N_{11-}}{N_{01-}+N_{11-}+N_{00+}+N_{10+}} \, .
\end{equation}

%%%\tcomment{These TPR and FPR are numbers. What is varied to produce a curve? The threshold $s_\mathrm{th}$? We fix that threshold using k-means.}

%%% \tnote{Is \href{https://developers.google.com/machine-learning/crash-course/classification/roc-and-auc}{area under the curve} relevant here?  In our case we fix the classification threshold parameter, i.e.\ how we split the predictions into positive and negative predictions, is fixed by k-means so we never scan through using other classifiers values. So do we care if we are better/worse for other classifier values?  Perhaps AUC is still another indication our method is better?}

%%%% ........................................................
%%%\subsubsection*{Precision}

Once the threshold $s_\mathrm{th}$ has been fixed, in our case using k-means \eqref{e:kmeans}, the \tdef{precision} score $S_{\mathrm{prec}}$ is defined as the number of times that we predict a link exists between node pairs in the later snapshot correctly (a true positive, $N_{11+}+N_{01+}$) divided by the number of times we predict a link, correctly (true positive) or incorrectly (false positive, $N_{11-}+N_{01-}$):
\bea
  S_{\mathrm{prec}}
  &=&
  \frac{ N_{11+}+ N_{01+} }{ N_{11+}+ N_{01+} + N_{11-}+ N_{01-} }  \, .
  \label{e:precdef}
\eea
A high precision score means we can trust that links predicted by the algorithm will exist.

We can set a baseline value for precision using a simple model which predicts a link in snapshot $\Gcal(s+1)$ exists for any given node pair with a probability given by the fraction of node pairs which have an link in snapshot $\Gcal(s)$, that is the density $\rho(s)$. In this case, the precision \eqref{e:precdef} is simply equal to the density in snapshot $\Gcal(s)$, $S_{\mathrm{prec,base}} =\rho(s)$, see Appendix F.

A very low precision score for baseline method indicates that the links existing between node pairs are not random. Some of the local measurement algorithms give lower scores than the baseline, suggesting that those types of local information are less likely to drive the evolution of the networks.

% --------------------------------------------------------------------------
\subsection*{Edge sampling}

We use the whole network to evaluate the performance, which is different from the sampling method used in~\figref{f:link_pred} to take account of the full nature of the data. In the following analysis, we predict the node pairs for all the possible node pairs of a whole network. We expect predictions can capture evolving characteristics of networks and the Mathematician network\st{s} change tiny fractions which are not sufficient enough to evaluate the predictions. Therefore, we replace the Mathematician network\st{s} with Hypertext network\st{s} (see Appendix C
% ~\appref{a:dataset}
for more details of this data set) in the following prediction analysis. Time data is sampled over different time intervals $\Delta t$. For Turkish Shareholder network\st{s}, the smallest time separation is $2$ years, and we consider four or more years too long for the evolution; for the Hypertext data, which records the short communications, we choose $\Delta t$ as $40$ and $60$ minutes; for the College Message and Email network\st{s}, we consider that $8$ hour to $1$ month communication frequency is reasonable.

%\newpage

% ***************************************************************************
\section*{Results}

\begin{figure}[htb!]
    \centering
    \includegraphics[scale=1]{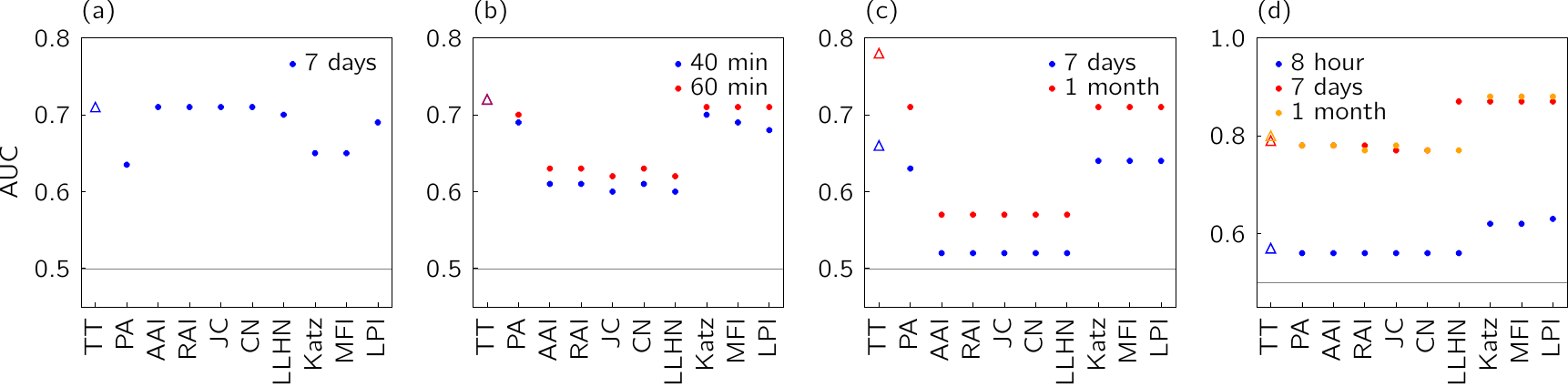}
    \caption[Link prediction AUC results]{The area under the curve (AUC) results for four temporal networks constructed from real data sets: (a) Turkish Shareholder network\st{s}, (b) Hypertext network\st{s}, (c) College Message network\st{s}, (d) Email network\st{s}. The results compare ten different methods of \tabref{t:linkprediction} including our TT algorithm denoted by the `triangle' symbol. For networks (b),(c) and (d), we show the results for different time scales (windows). See \tabref{t:linkprediction} for abbreviations used in indicate link prediction methods.
	}
    \label{f:predictionauc}
\end{figure}

\Figref{f:predictionauc} shows a comparison of AUC for the ten algorithms when we apply them to node pairs sampled uniformly at random from all possible node pairs. The Triplet Transition (TT) approach defined here is the left-most triangular point in each figure. In the Hypertext network\st{s} (b) and the College Message network\st{s} (c), the Triplet Transition  method has the highest AUC though the PA, Katz, MFI and LPI algorithms perform almost as well (see \tabref{t:linkprediction} for abbreviations).
For the Turkish \hl{S}hareholder network (a) the AUC of our Triplet Transition  method is again the highest though with a similar AUC value for various algorithms.  Note that PA, Katz, and MFI are now weaker. Only for the Email network (d) is our Triplet Transition  outperformed, in this case by the Katz, MFI and LPI algorithms.

\begin{table}[!ht]
    \centering
    \resizebox{\columnwidth}{!}{%
    \begin{tabular}{l|l||l|l||l|l||l|l||l|l||c}
      \multirow{3}{*}{Type} & Algorithm  & \multicolumn{2}{c||}{Shareholder} & \multicolumn{2}{c||}{Hypertext} & \multicolumn{2}{c||}{CollegeMsg} & \multicolumn{2}{c||}{Email} &  Avg.
            \\
      & (Time scale) & \multicolumn{2}{c||}{(2 years)}     & \multicolumn{2}{c||}{(1 h)}              & \multicolumn{2}{c||}{(1 mon)}               & \multicolumn{2}{c||}{(1 mon)}                &      Rank
      \\
      & $[\overline{f_\Ecal (s)}]$ & \multicolumn{2}{c||}{$[0.63]$}        & \multicolumn{2}{c||}{$[0.52]$}            & \multicolumn{2}{c||}{$[0.93]$}                & \multicolumn{2}{c||}{$[0.50]$}                 &
      \\ \hline\hline
     \multirow{4}{*}{Global} & \textit{Triplet Transition}  & $\mathbf{0.71}$  & $1$ & \multicolumn{1}{l|}{$\mathbf{0.72\,(7)}$}   & $1$    & $\mathbf{0.77\,(6)}$   & $1$  & $0.80\,(2)$   & $4$  & $1.8$    \\
      &\textit{Katz} ($\beta=0.01$)     & $0.65$            & $8$ & \multicolumn{1}{l|}{$\mathbf{0.71\,(10)}$}   & $1$    & $0.69\,(8)$& $2$ & $\mathbf{0.88\,(2)}$ & $1$  & $3.0$ \\
    &\textit{Matrix Forest Index}      & $0.65$    & $8$      & \multicolumn{1}{l|}{$\mathbf{0.71\,(9)}$}   & $1$    & $0.69\,(8)$  & $2$  & $\mathbf{0.88\,(2)}$     & $1$  & $3.0$ \\
    & \textit{Local Path Index}       & $0.69$   & $7$   & \multicolumn{1}{l|}{$0.69\,(9)$}   & $5$  & $0.69\,(6)$   & $2$    & $\mathbf{0.88\,(2)}$    & $1$       & $3.8$  \\ \hline
    \multirow{6}{*}{Local}   &\textit{Resource Allocating Index}        & $\mathbf{0.71}$            & $1$            & \multicolumn{1}{l|}{$0.63\,(8)$}   & $6$    & $0.58\,(3)$ & $6$  & $0.78\,(2)$ & $5$                 & $4.5$ \\
    &  \textit{Adar-Academic Index}       & $\mathbf{0.71}$            & $1$            & \multicolumn{1}{l|}{$0.63\,(8)$}   & $6$    & $0.58\,(3)$   & $6$  & $0.78\,(2)$   & $5$   & $4.5$  \\
    &  \textit{Common Neighbour   }        & $\mathbf{0.71}$    & $1$   & \multicolumn{1}{l|}{$0.63\,(8)$}   & $6$    & $0.58\,(3)$    & $6$   & $0.77\,(2)$   & $8$  & $5.3$  \\
    & \textit{Jaccard Coefficient}           & $\mathbf{0.71}$    & $1$  & \multicolumn{1}{l|}{$0.62\,(8)$} & $9$    & $0.58\,(2)$  & $6$    & $0.77\,(2)$   & $8$    & $6.0$ \\
    & \textit{Preferential Attachment}           & $0.63$            & $10$  & \multicolumn{1}{l|}{$0.70\,(7)$} & $4$ & $0.69\,(6)$  & $2$   & $0.78\,(1)$  & $5$  & $5.3$  \\
    & \textit{Local Leicht-Holme-Newman}      & $0.70$   & $6$   & \multicolumn{1}{l|}{$0.62\,(7)$}   & $9$   & $0.57\,(2)$  & $10$   & $0.77\,(2)$   & $8$    & $8.3$ \\ \hline
    & \textit{Baseline}     & $0.5$  & $11$   & \multicolumn{1}{l|}{$0.5$}   & $11$   & $0.5$  & $11$  & $0.5$ & $11$  & $11.0$
   \end{tabular}
        }% end of resizebox
    \caption{Summary of AUC-ROC performance, $\Delta t$ is taken by selecting the highest AUC-ROC. The AUC scores for each network are in the left-hand column, the rank of the score in the right-hand column. The errors quoted on the AUC scores are standard deviations from all the computed network snapshots of the selected time scale except for \hl{Turkish S}hareholder network\st{s} where only two predictions are made. We highlight the highest result and any whose result is within the error quoted on the largest result. The baseline method for AUC is the random clustering of node pairs into two groups, and the AUC is $0.5$ which means it is like `tossing a coin'. The Type column indicates if the link prediction method probes only ultra scales (path length two) or global scales (infinte path lengths) }
    \label{t:predictionauc}
\end{table}

% ........................................................
%\subsubsection*{Precision results}

\begin{figure}[htb!]
	\centering
	\includegraphics[scale=1]{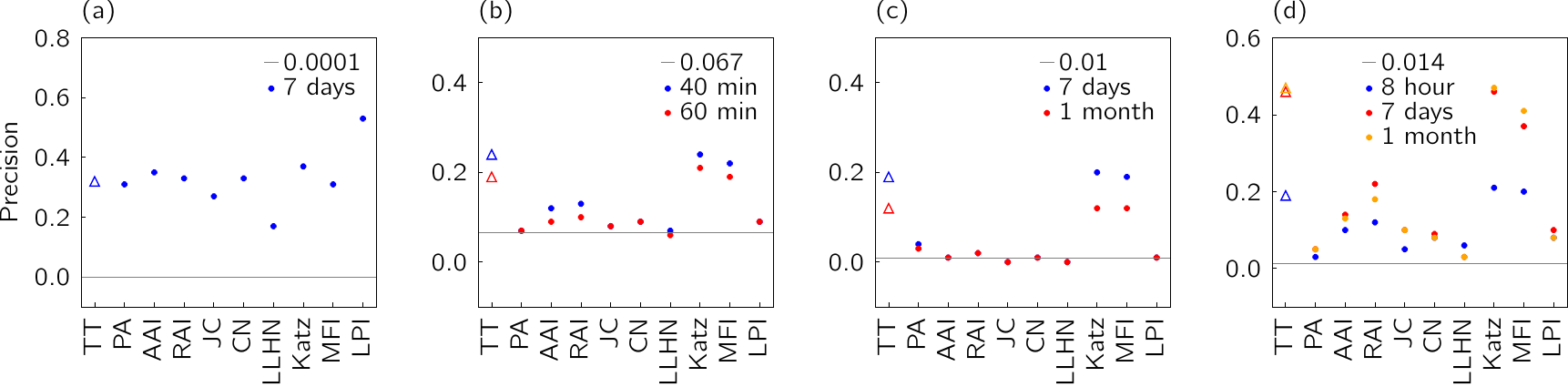}
	\caption[Link prediction precision results]{Precision scores for the link prediction results for ten algorithms applied to four temporal networks constructed from real data sets: (a) Turkish Shareholder network\st{s}, (b) Hypertext network\st{s}, (c) College Message network\st{s}, (d) Email network\st{s}. Results for the Triad Transition algorithm based on transition matrix denoted by the triangle symbol $\bigtriangleup$. For the networks in (b), (c) and (d), we also show the results for different time scales (window). See \tabref{t:linkprediction} for abbreviations used in indicate link prediction methods.
	}
	\label{f:predictionprec}
\end{figure}

\Figref{f:predictionprec} shows that there are some clear patterns in our results for precision.  First, each link prediction method has a similar performance relative to the other methods on three of the networks: the Hypertext network (b), the College Message \hl{n}etwork (c), and the Email network (d).  On these three data sets, three algorithms are consistently better than the others though with similar performances relative to one another: our Triplet Transition (TT) method, the Katz method~\cite{LK07,LJZ09a,AMA20}, and the Matrix Forest Index~\cite{CS97} (MFI) method.  All of these are probing non-local information in the networks, suggesting this is necessary to understand the time evolution of these networks.

\begin{table}[htb!]
	\resizebox{\columnwidth}{!}{%
		\begin{tabular}{l|l||l|l||l|l||l|l||l|l||c}
			\multirow{3}{*}{Type} & Algorithm  & \multicolumn{2}{c||}{Shareholder} & \multicolumn{2}{c||}{Hypertext} & \multicolumn{2}{c||}{CollegeMsg} & \multicolumn{2}{c||}{Email} &  Avg.
			\\
			& (Time scale) & \multicolumn{2}{c||}{(2 years)}     & \multicolumn{2}{c||}{(1h)}              & \multicolumn{2}{c||}{(1 mon)}               & \multicolumn{2}{c||}{(1 mon)}                &      Rank \\
			& $[\overline{f_\Ecal (s)}]$ & \multicolumn{2}{c||}{[$0.63$]}        & \multicolumn{2}{c||}{[$0.52$]}            & \multicolumn{2}{c||}{[$0.93$]}                & \multicolumn{2}{c||}{[$0.50$]}                 &      \\ \hline\hline
			\multirow{4}{*}{Global}
			& \textit{Triplet Transition}   & $0.27$  & $5$ & \multicolumn{1}{l|}{$0.19\,(9)$}   & 2    & $0.11\,(4)$                & $2$  & $\mathbf{0.47\,(6)}$  & $1$  & $2.5$    \\
			& \textit{Katz} ($\beta=0.01$)     & $0.25$            & $8$            & \multicolumn{1}{l|}{$\mathbf{0.22\,(8)}$}   & $1$    & $0.11\,(4)$& $2$ & $\mathbf{0.47\,(6)}$ & $1$  & $3.0$ \\
			&\textit{Matrix Forest Index}        & $0.19$    & $9$      & \multicolumn{1}{l|}{$0.19\,(9)$}   & $2$    & $\mathbf{0.12\,(3)}$  & $1$  & $0.4\,(5)$     & $3$  & $3.8$ \\
			& \textit{Local Path Index}         & $\mathbf{0.36}$   & $1$   & \multicolumn{1}{l|}{$0.09\,(4)$}   & $5$  & $0.009\,(2)$   & $7$    & $0.08\,(3)$    & $7$       & $5.0$  \\ \hline
			\multirow{6}{*}{Local}
			& \textit{Resource Allocating Index}          & $0.28$            & $4$            & \multicolumn{1}{l|}{$0.10\,(6)$}   & $4$    & $0.019\,(4)$                & $5.0$                & $0.18\,(7)$ & $4$                 & $4.3$ \\
			& \textit{Adar-Academic Index}       & $0.33$            & $2$            & \multicolumn{1}{l|}{$0.09\,(5)$}   & $5$    & $0.008\,(2)$   & $8$  & $0.13\,(5)$   & $5$   & $5.0$  \\
			&  \textit{Common Neighbour   }         & $0.31$    & $3$   & \multicolumn{1}{l|}{$0.09\,(5)$}   & $5$    & $0.008\,(2)$    & $8$   & $0.08\,(1)$   & $7$  & $5.8$  \\
			&  \textit{Jaccard Coefficient}         & $0.25$    & $7$  & \multicolumn{1}{l|}{$0.08\,(6)$} & $8$    & $0.0010\,(9)$                 & $10$               & $0.10\,(4)$   & $6.0$    & $7.8$ \\
			&  \textit{Preferential Attachment}             & $0.27$            & $6$  & \multicolumn{1}{l|}{$0.07\,(6)$} & $9$ & $0.03\,(1)$  & $4$   & $0.059\,(6)$  & $9$  & $7.0$  \\
			& \textit{Local Leicht-Holme-Newman}       & $0.09$   & $10$   & \multicolumn{1}{l|}{$0.06\,(5)$}   & $11$   & $0.0001\,(2)$  & $11$   & $0.03\,(1)$   & $10$                & $10.5$ \\ \hline
			& \textit{Baseline}   & $0.0001$  & $11$   & \multicolumn{1}{l|}{$0.067$}   & $10$   & $0.010$  & $6$  & $0.014$ & $11$  & $9.5$
		\end{tabular}
	}
	\caption{Summary of the average precision scores (on the left) and their ranks (on the right) for selected $\Delta t$ based on k-means clustering. The errors quoted are standard deviations from all the computed network snapshots of the selected time scale, except for the\hl{Turkish} \hl{S}hareholder network\st{s} which have one prediction for three snapshots.}
	\label{t:predictionprec}
\end{table}

Overall we see some similarity in these AUC results, summarised in \tabref{t:predictionauc} as we saw for precision in \tabref{t:predictionprec}. The same three algorithms, the Triplet Transition (TT) methods, the Katz method, and the MFI method, have a similar high performance on the same three networks, the Hypertext network, the College Message \hl{n}etwork and the Email network.  For these three networks, we might also pick out the Preferential Attachment (PA) algorithm. However, now the AUC values for the \hl{Turkish S}hareholder network show that our Triplet Transition method continues to perform well, unlike for the precision measurements. The Katz, Matrix Forest Index and Preferential Attachment  methods are, though, in a weaker group of algorithms as measured by their AUC performance on the\hl{Turkish S}hareholder network.

For three of these networks, we also measure AUC over snapshots of these networks defined over different time scales.
The comparison among methods rarely changes. However, the performance of most algorithms is altered by the size of the time window chosen in some cases, reflecting inherent timescales in the different systems. The most noticeable is that for the Email network, there is a large difference between windows of eight hours and one week but not so much change between a week and one month.  The gaps may suggest that if a person is going to produce an email, perhaps following up an email request of bringing a third person into the conversation, that new email is done often on the scale of a few days not always on the scale of a few hours.

The Triplet Transition proposed here is used to predict a link between two nodes by considering these alongside a third node which can be in any position in the network. In this way, this third node not only captures the higher-order interactions but also enables the method to encode both local and non-local information about the link of interest. We find that our Triplet Transition method and the two other global methods used here, Katz Index method~\cite{LK07,LJZ09a,AMA20} and the Matrix Forest Index~\cite{CS97} are generally the best for most of networks we studied, in particular for the Hypertext network\st{s}, the College Message network\st{s} and the Email network\st{s}. As the most successful methods here perform better than other approaches based on local measures, this shows that in most systems the pattern of connections depends on the broader structure of interactions. This dependence of the behaviour of systems on structure beyond nearest neighbours is the crucial motivation for using the language of networks rather than just looking at the statistics of pairs \cite{BRMW13}.  For instance, the Katz method counts the number of paths between each pair of nodes, probing all paths though giving less weight to longer paths, so nearest neighbours contribute the most.

The results for the{Turkish} Shareholder network\st{s} were a little different. In this case, predictions based on local measurements (paths of length two), the semi-local Local Path Index method (LPI), as well as our non-local triplet transition method outperformed the other global methods in terms of AUC, see \tabref{t:predictionauc}.  The global methods also perform poorly on precision, see \tabref{t:predictionprec}. An algorithm with low precision and high AUC, such as Jaccard Coefficient (JC),  is predicting the disconnected pairs well.

% ***************************************************************************
\section*{Discussion}

In this paper, we considered the temporal evolution of networks by looking at a sequence of snapshots of each network.  The network $\Gcal(s)$ defined for snapshot $s$ contains all the links present between nodes for a certain period, $\Delta t$. Each snapshot $s$ covers the period $\Delta t$ immediately following the latest time included in the previous snapshot. In some cases, we have looked at the effect of changing the size of the temporal windows $\Delta t$ on our results.
The main tool we use to study the network evolution is the transition matrix $\Tsf$ of equation~\eqref{e:Tdef} which are derived from the evolution of three-node combinations in a network from one temporal snapshot $\Gcal(s-1)$ to the next $\Gcal(s)$.  These transition matrices are obtained from real data by counting the different states of three nodes found in consecutive network snapshots.

To decode the higher-order interaction patterns of network dynamics, we fitted a pairwise interaction model to the real data and computed the transition matrices from the fitted parameters. By comparing the actual transition matrices found numerically with those predicted using a simple pairwise model, as shown in~\figref{f:tm_real}
% \hl{Figure 3}
, we demonstrated that higher-order interactions are needed to understand the evolution of networks.
For example, in the Email network, derived from the emails within an EU Institution, if three nodes are connected in the previous temporal snapshot, they are less likely to be disconnected in the following snapshot.

To show that we must look beyond the interactions of neighbours, we then designed a link prediction algorithm based on the transition matrix $\Tsf$ of equation~\eqref{e:Tdef}, our Triplet Transition (TT) method. We compared this method with nine other link prediction methods as well as to simple baseline measures. What we found was that on a range of different temporal networks, the Triplet Transition method was as good as two methods based on non-local (global) information in the network, namely, the Katz Index method~\cite{LK07,LJZ09a,AMA20} and the Matrix Forest Index (MFI) method~\cite{CS97}. While not always the best on every network or every measure, these three global methods were usually better than the other methods we studied, all of which used information on paths of length two in the network. Intriguingly, the one other method that used paths of length three, Local Path Index~\cite{ZLZ09,LJZ09a}  (LPI), often performed well too though rarely as well as the top three global methods.
%In the Hypertext networks and the College message networks, our TT has the highest AUC and precision values. In the shareholder networks, the AUC of TT is the highest, and its precision is the second-highest among all the algorithms; in the Email networks,  TT's AUC has the second-highest value, and its precision is the highest.

Since the most successful methods in our tests were those that access non-local information, it seems that such information is essential in the evolution of most networks and therefore, it is important to include this in network measures. However, including information from the whole network is numerically intensive and, for any reasonably sized network, the evolution of a link is unlikely to depend directly on what is happening a long way from that link. One reason why our Transition Triplet method works well is that it does not emphasise the vast majority of the network.  Most of the information in the transition matrix network is based on neighbours of one or other of the link of interest.  For large networks, we use sampling to add the necessary global information into the transition matrices. The Katz index method does include information from all scales but suppresses contributions from more distant parts of the network. The success of the transition triplet approach suggests that there is no need to access all of the global information of a network in order to know what is going on locally. Notably, the overall better performance of the Triplet Transition method reveals that the information of different higher-order interaction patterns can help understand and predict different dynamics of networks.

There is also another reason why our Triplet Transition method may work better than the local methods we look at, and that is because our method is also probing a longer time scale as well as a longer spatial scale.  That is we use \emph{two} snapshots, $\Gcal(s-1)$ and $\Gcal(s)$ in order to create the transition matrix $\Tsf$ which in turn we use for the predictions in snapshot $\Gcal(s+1)$. All the other methods used here are based on information from one snapshot $\Gcal(s)$. So again, the success of the Triplet Transition method points to correlations over short but non-trivial time scales as being important in understanding network evolution.  In our case, the dependence of results on the time intervals can be seen in the effect of $\Delta t$ used to define the transition matrices are important. For different data sets, we found different $\Delta t$ gave optimal results which of course reflects inherently different timescales in the processes encoded by our different data sets. So another conclusion is that higher order effects in terms of both time and space are needed to understand the evolution of temporal networks and to make effective predictions for links. The inclusion of higher-order order effects in terms of time remains undeveloped relative to the work on higher-order spatial network features.

\hl{In our approach we have focused on changes in the edges and ignored changes in the set of nodes. Our method can include nodes which are not connected in some or even all our snapshots provided these nodes are known. However, in many cases the data sets may only record interactions, our edges, and our set of nodes is only inferred from that.  As a result we often have no information about such (totally isolated) nodes.  Suppose we are given a list of emails, our edges, sent at a given time between email accounts, our nodes. We cannot distinguish accounts which are dormant from those that are deleted or added to the system without additional information. Should we have such data on node birth and death, then in some cases it might be an important process to include, but it is not one we address in our approach.}

Currently constructing optimal higher order models is a timely topic~\cite{RMI2019}. Our method can be generalised to include even higher order interactions, such as quadruplet and so on. The method can be used to indicate higher order evolutionary mechanisms in a network and suggest what is the most likely order of interaction.  Our work shows that studying the evolution of small graphs over short time periods can reveal important information and predictions regarding network evolution.

% *********************************************

\clearpage
% \input{tripletsbibliography}
% Bibliography for triplet.tex

% \providecommand{\doinumber}[1]{\href{http://doi.org/#1}\texttt{#1}}
% \providecommand{\journal}[1]{#1}
% \providecommand{\volume}[1]{\textbf{#1}}
% \providecommand{\number}[1]{}
% \providecommand{\pages}[1]{#1}
% \providecommand{\year}[1]{#1}

% *****************************************************************
%
%   BIBLIOGRAPHY
%

\section*{Acknowledgement}

This research is supported by the Centre for Complexity science group at Imperial College London. We thank Nanxin Wei, from Imperial College London and Junming Huang, from Princeton University for developing early ideas of this paper. We thank Hardik Rajpal and Fernando Rosas for valuable discussion of the triplets in the graph. We also appreciate another colleague Hayato Goto for valuable interoperation of machine learning methods and its application.

\section*{Author contributions statement}
Q.Y. and B.C. conducted the numerical simulations and data analysis. Q.Y., B.C., T.E. and K.C. analysed and interpreted the results, and wrote the manuscript.

\clearpage

% This file should be read in to prove the text and figures for the appendix/Supplemntary Information
% 18/5/21

% ************************************************************
\section{Transition matrix estimation}\label{a:tf_estimation}

\begin{figure}[!h]
    \centering
    \includegraphics[scale=1]{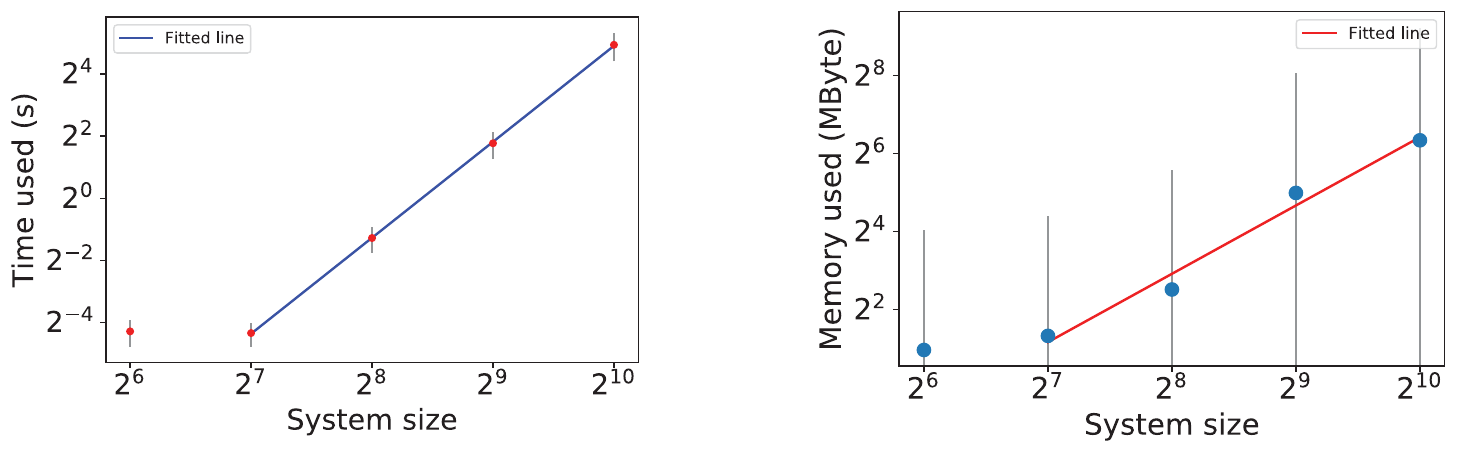}
    \caption{Time and memory needed for generating all three node graphlet combinations against a different number of nodes. $N=2^n$ $n=6,7,8,9,10$. For each $n$, the time and memory for each run are measured and the error bars are corresponding to standard deviation. The time used scales as expected order (the slope of the fitted line is $3.09\pm 0.02$) of the system size. The slope of the fitted line for the memory used is $1.75 \pm 0.02$.}
    \label{fig:time_mem}
\end{figure}

The memory needed for the calculations in our Triplet Transition (TT) method scales with the number of combination of triplet in network, that is  $\binom{N}{3} = N(N-1)(N-2) / 6  \sim O(N^3)$ and this can be seen in \figref{fig:time_mem}.
%%%For example: for a graph of $1000$ nodes, the combination of two nodes is around $(10^3)^2 = 10^6$ while the combination of the three nodes is around $(10^3)^3 = 10^9$.
%While the number of nodes in a graph increase, the extra memory grows exponentially for computing the interactions among three nodes.

To ensure the number of triplets is sufficiently large enough to estimate the transition matrix $\Tsfhat$, we use all three-node combinations if the number of nodes in the system is smaller than $10^3$; otherwise, we sample $10^5$ triplets chosen uniformly at random from the set of all three-node combinations. If the calculated $\Tsfhat$ is stable, we choose this sample number for the following analysis. Otherwise, we continue to add sample a further sample of $10^5$ triplets until the $\Tsfhat$ is stable.

% *******************************************************
\section{Simple Null Models}\label{a:simple_model}

The simplest null model for the evolution of the network is one in which we assign a probability $p$ that in each time step, a pair of nodes changes from disconnected to connect with probability $p$. In contrast, a connected pair becomes disconnected with probability $q$. We can then write down the transition matrix in this Pairwise Null model $\Tsfpw(p,q)$ (often abbreviated to $\Tsfpw$) for our three-node combinations in terms of the set of four states $\Triadset^4=\{m_0,m_1,m_2,m_3\}$ where $m_i$ is the configuration in which three nodes have $i$ links between them. Namely
\begin{equation} \label{eq:tm_pairwise}
        \Tsfpw (p,q)
       \! = \!\!
        \begin{pmatrix}
        (1\!\!-\!p)^3 & 3p(1\!\!-\!p)^2 &3p^2(1\!\!-\!p) & p^3 \\
        q(1\!\!-\!p)^2 & (1\!\!-\!q)(1\!\!-\!p)^2+2qp(1\!\!-\!p) & 2(1\!\!-\!q)p(1\!\!-\!p)+qp^2 & (1\!\!-\!q)p^2 \\
        q^2(1\!\!-\!p) & 2(1\!\!-\!q)q(1\!-\!p)\!+\!q^2p & (1\!\!-\!\!q)^2(1\!-\!p)+2qp(1\!-\!q) & (1\!-\!q)^2p \\
        q^3 & 3(1\!-\!q)q^2 & 3q(1\!-\!q)^2 & (1\!-\!q)^3 \\
        \end{pmatrix}.
\end{equation}
Here $\Tsfpw_{ij}$ is the probability that three nodes connected with $i$ nodes, in state $m_i$, evolves in the next time step to a configuration $m_j$ with $j$ edges between those three nodes.
For instance, $\Tsfpw_{03}$ denotes the probability of evolving from three disconnected nodes $m_0$ to state $m_3$ of three fully connected nodes. The addition of an edge for each of the three edges gives a single factor of $p$, so $\Tsfpw_{03}=p^3$ overall. When looking at the transition from a single edge to two edges in a triplet, the entry $\Tsfpw_{12}$, we can do this in two ways. First, we can add one new edge with probability $p$, keeping the other pair of unconnected nodes in that state, probability (1-p), and keeping the original single connected pair in that state with probability $(1-q)$. There are two ways of choosing where to add the extra edge, so we have a factor of $2p(1-p)(1-q)$.  However, we could also remove the existing edge and add two new edges between the other pairs of edges edge in one of two positions, giving the second factor of $p^2q$ seen in the entry for $\Tsfpw_{12}$ in  \eqref{eq:tm_pairwise}. We can check that the rows sum to one, $1=\sum_{j} \Tsfpw_{ij}$ since we conserve the total number of triplets in our models.

%%%If we want to know how much closed triads($m_3$) evolve to a unconnected three node pair($m_0$), i.e.,entry $\Tsfpw_{30}$. The process of $\Triad_{s+1} = m_0$ transits from $\Triad_{s} = m_3$ can be treated destroying all the $3$ edges independently in $m_3$ at $t$. The probability of destroying an existing edge is $q$, therefore, destroying $3$ edges idependently will be $q^{3}$.
%%%(Bingsheng updated this sentences. Previous comment: Qing, I do not know how to edit the sentence to make it clearer.) Furthermore, since we do not specify which two nodes are connected by this edge in triads, there are $3$ degenerate states(combinations) and link can appear at any positions between three nodes, therefore, the probability of $m_1$ and $m_2$ multiply by $3$. For $m_{0}$ and $m_{3}$ cases, there is no degenerate state and therefore, we do not need to multiply any pre-factors.

Our second null model, our ``edge swap model'', we swap the ends of a pair of edges so edges $(u,v)$ and $(w,x)$ in snapshot $s$ are removed and are replaced by edges $(u,x)$ and $(w,v)$ in the next snapshot. This preserves the degree of every node. For each update from $\Gcal(s)$ to $\Gcal(s+1)$ we update 20$\%$ edges.

The final model, our ``random walk model'', is also an edge rewiring model but it is based on higher-order structures as we use random walks to select the new edges.
The model starts with an Erd\H{o}s-R\'{e}nyi graph. The initial node for a random walker, say $u$, is chosen uniformly at random from the set of nodes. Then one of the edges from $u$, say the edge to a node $y$ is chosen uniformly from the set of neighbours. Finally, three non-backtracking steps are made on the network starting from $u$ and ending at a node $x$, in which edges are always chosen uniformly from those available excluding any edge used in the previous step of the random walk. The existing edge $(u,y)$ is removed and replaced by a new edge $(u,x)$.
The final graph is then created through a projection where nodes are connected if they share a common neighbour, that is if directed edges from $u$ to $v$ and $w$ to $v$ exists, then the projected graph has an undirected link between $u$ and $v$.  This rewiring and projection procedure maintains the number of edges and nodes in the original graph but not in the projected graph.

To produce the time evolution, we rewire 20\% of the edges using this rewiring procedure and then use this new network as the next snapshot.
In our context, our random walk model is used to produce test networks with local correlations between nodes.  While motivated by real-world examples and building on existing experience with the model \cite{QY}, here it is used as a toy model to illustrate the approach.

\begin{figure}[htb!]
    \centering
    \includegraphics[width=0.9\textwidth]{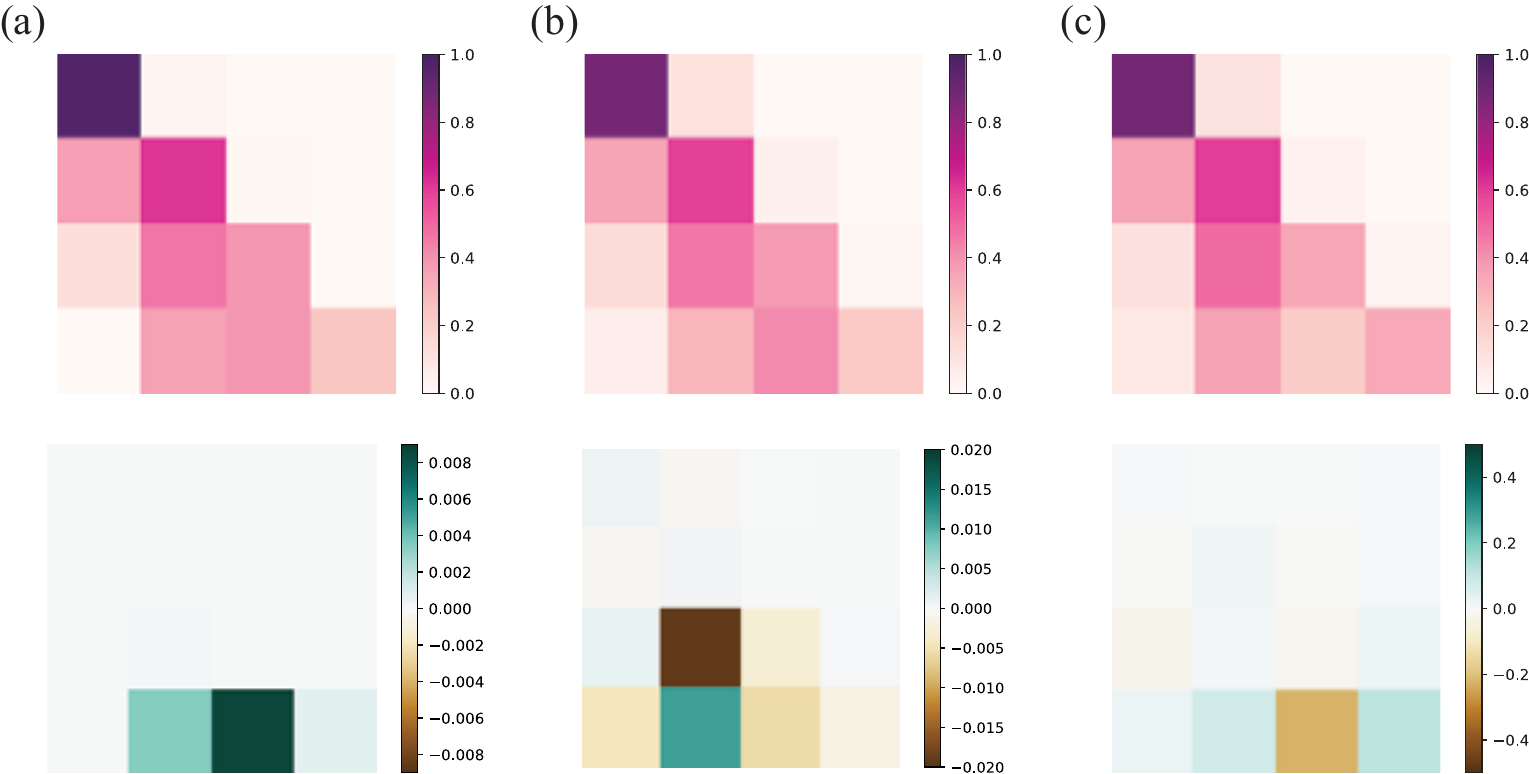}
    \vspace*{5mm}
    \caption{The triplet transition matrix $\Tsfhat$ estimated from the artificial data. Each of the six four-by-four heat maps shows values where the rows represent one initial triplet graphlet $i$ and the columns representing the final graphlet $j$ using our unlabelled graphlet $\Mcal_4$ set of triplets to make the visualisations manageable.
    The top three heat maps show the average value of entries $\texpect{\Tsfhat_{ij}}$.
    The bottom row of three heat maps give the values of the difference matrix $\texpect{\Delta \Tsfhat} = \Tsfhatpw - \Tsfhat$ of \eqref{a:deltaTsfhatpwdef}
    in which the numerical data is compared to the analytical form predicted from the simple pairwise model of  \eqref{eq:tm_pairwise}.
    The scales for the colours of the first two heat maps on the bottom row are much smaller (a factor of twenty or more)
    than for the results for $\texpect{\Delta \Tsfhat} = \texpect{\Tsfhatpw} - \texpect{\Tsfhat}$ in the  random walk model shown in the bottom right corner.
    %he bottom three heat maps give the values of the difference matrix $\texpect{\Delta \Tsfhat} = \Tsfhatpw - \Tsfhat$ of  (8)
    % \eqref{ae:deltaTsfhatpwdef}
    Any large entries in these lower rows of heatmaps indicate higher-order effects not present in our simple pairwise model of $\Tsfhatpw$. The three columns of heat maps show results for the three different artificial temporal networks created numerically using the stochastic models defined in \secref{a:simple_model}: (a) our ``pairwise model'', (b) our ``edge swap model'', and (c) our ``random walk model''.
    Only the networks formed with random walks show significant higher-order effects.
    }
    \label{fig:tm_toy}
\end{figure}

We can use these three simple models to generate artificial temporal networks to test our approach. The results in terms of the transition matrix $\Tsfhat$ derived from the artificial networks are shown in \figref{fig:tm_toy}. To show the deviation from the our pairwise interaction null model,  we look at the difference $\Delta \Tsfhat(s)$ between actual results for the average of $\Tsfhat(s)$ and those predicted in the null model $\Tsfhatpw(s)$, so
\begin{equation}
 \Delta \Tsfhat(s) = \Tsfhat(s) - \Tsfhatpw(s) \, .
 \label{a:deltaTsfhatpwdef}
\end{equation}

The behaviour of our simple ``pairwise model'' should be completely captured by the reference transition matrix $\Tsfhatpw(s)$ and, as expected, the numerical results shown in \subfigref{fig:tm_toy}{a} show no significant difference between numerical data $\Tsfhatpw(s)$ and theoretical $\Tsfpw$.

For \subfigref{fig:tm_toy}{b}, we use the artificial networks generated by the ``edge swap model''.  While this involves two pairs of edges, so in principle is a higher-order model, in a sparse graph, the four nodes selected by a pair of randomly selected edges are unlikely to be linked by other edges.  So in practice, in terms of the triplet graphlets, this model behaves much like the pairwise model and shows little difference from that model.

It is only with the networks generated using our three-step random walk that we see significant differences between the data and the pairwise model. This is to be expected as higher-order processes were used to create the numerical networks.

% ************************************************************
\section{Data Sets}\label{a:dataset}

In this project, we use several different datasets to produce temporal networks with different resolutions. The resolution of a network is the time interval used to create each snapshot $\Gcal(s)$ of our network. This can be done in two ways, depending on the context. In either case, the resolution can be `seconds', `hours', `days' or `months' and should be chosen to suit the context.

In the first type of temporal data, the data capture interactions between pairs of nodes which occur briefly on the time scale of the network resolution. A list of the times of phone calls between members of a social network would be an example of this type of data. In this case, each edge in a single snapshot indicates that an event linking the two nodes occurred during the time interval.

The second approach is where the pairwise interactions recorded in the data typically last for much longer than the resolution, but they do change slowly over time.  An example of this would be hyperlinks between webpages. In this second case, the snapshots are the network at one instant in time, and the interval is now the time between these snapshots.

We use five different data sets which are as follows.
\begin{itemize}
    \item \texttt{Turkish Shareholder Network} (\texttt{Shareholder}).

    The nodes are shareholders in Turkish companies. The shareholders are linked if they both hold shares in the same company during the time interval associated with the snapshot \cite{QY}.

    %%%     have a common sharholding during Turkish shareholder network: This is the projection of the shareholder and company network. The nodes represent shareholders and a link means two connected shareholders invested in at least one common company \cite{QY}.

    \item \texttt{Wikipedia Mathematician} (\texttt{WikiMath}). Each biographical Wikipedia page of an individual mathematicians forms a node. If a hyperlink links two biographies (in either or both directions), a link is present in the network. The edges are edited by users and are both added and removed over time. Each snapshot represents the state of these webpages at one moment in time. The data is taken at one point in three different years, 2013,2017, and 2018, so the intervals are not constant in this case. See the paper~\cite{BZT19} for further details on this dataset.

    %Social network. references will be added later):

\item \texttt{College Message} (\texttt{CollegeMsg}).
%This is a communication network.
The nodes are students, and an edge in a snapshot indicates that the students exchanged a message within the interval associated with that snapshot. The data was collected over a seven month period in 2004, see \cite{OP09,POC09,SNAP} for more details.

    %%%\tnote{The Bibtex for these references is commented out here in this file.}

    \item \texttt{Email} (\texttt{Email}).
    %This is a communication network.
    This is derived from the emails at a large European research institution sent between October 2003 and May 2005 (18 months). Each node corresponds to an email address. An edge in a given snapshot indicates that an email was sent between the nodes in the time interval corresponding to that snapshot \cite{SNAP,LKF07,EUEmail}.

    %\tnote{The Bibtex for these references is commented out here in this file.}

    \item \texttt{Hypertext} (\texttt{Hypertext}).
    This is the network of face-to-face contacts of the attendees of the Association of Computing Machinery (ACM) Hypertext 2009 conference. In the network, a node represents a conference visitor, and an edge represents a face-to-face contact that was active for at least 20 seconds \cite{KONECT,Hypertext}.

    %\tnote{I could not confirm the original source of this data nor can I find KONECT.}
    % Multiple edges denote multiple contacts. Each edge is annotated with the time at which the contact took place.

%%%    %%% Communication network (Email-Institution, College Message)
%%%    This is the email communication network of a large, undisclosed European institution. Nodes represent individual persons. Edges between two persons are directed and denote that at least one email has been sent from one person to the other. All edges are simple: Even if a person has sent multiple emails to another person, the two persons will be connected only by a single edge in that direction. Spam emails have been removed from the dataset. \cite{EUEmail}.

%%%    \item Artificial network with known mechanism
%%%
%%%    Projection graph of random-walk on bipartite network: There are two types of the nodes in the projected network. In each snapshot of network, graph is constantly evolving and a fraction of the edges are removed and rewired in every snapshot of the network. There are two mechanism of rewiring in the network, one type of nodes only like to connect to the nodes with two or more edges on a bipartite network and vice versa for other node. We only use the projected network for prediction.
\end{itemize}

We provide a summary of the graph statistics in \tabref{tab:networksummary}. Some further information on the temporal characteristics of some of the data sets is given in Figures
\ref{fig:email8h} --
%(\ref{fig:email7days}, (\ref{fig:email1month}, (\ref{fig:college7day} and
\ref{fig:college1mon}.

\begin{table}[htb!]
\centering \footnotesize
%\newlength{\timwidth}
\settowidth{\timwidth}{7 days, 1 month}
\begin{tabular}{rl||c|c|c|c|c}  %\overline{\overline{}}
                 Dataset & (Abbreviation)         & Nodes  & Edges & Time Period          & Resolutions              & Source \\
                         &                        & ($N$)  & ($E$) & ($T$)                & ($\Delta t$)             &         \\[2pt]  \hline \hline
     Turkish Shareholder & (\texttt{Shareholder}) &  39901 & 68017 & 2010,2012,2014,2016  & 2 years                  & \cite{QY}     \\[6pt] \hline
\parbox[t]{\timwidth}{\centering Wikipedia \\ Mathematician\st{s}}
                         & (\texttt{WikiMath})    &   6049 & 36315 & 2013,2017,2018       & 1 and 4 years            & \cite{BZT19}  \\[12pt] \hline
         College Message & (\texttt{CollegeMsg})  &   1899 & 20296 & 6 months, 2004       & 7 days, 1 month          & \cite{OP09,POC09,SNAP}   \\[6pt]  \hline
       Institution Email & (\texttt{Email})       &    986 & 24929 & 1 year, 1970-1971    &
       \parbox[t]{\timwidth}{\centering 8 hours, 7 days,\\1 month} & \cite{SNAP,LKF07,EUEmail} \\[12pt] \hline
               Hypertext & (\texttt{Hypertext})   &    113 &  5246 & 3 days in 2009 & 40,60 min                & \cite{KONECT,Hypertext}  \\
%Hep-Ph  & 28,093 & 4,596,803 & 1993-2003 (yr) & 1 day                \\
\end{tabular}
\caption{The detailed information of graph statistics.}
\label{tab:networksummary}
\end{table}

\begin{figure}[h!]
\centering
\includegraphics[width=0.9\textwidth]{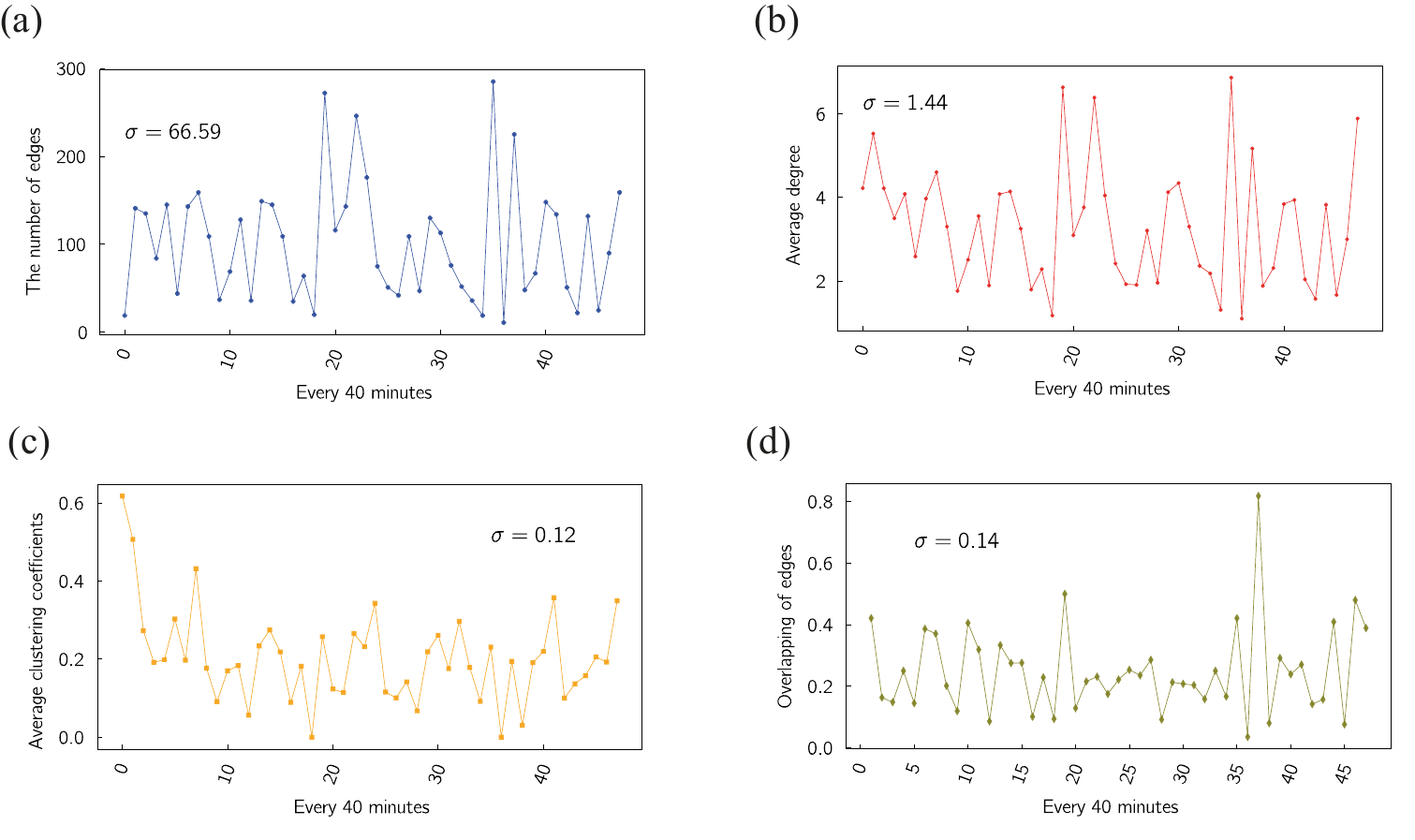}
\caption{Temporal evolution of graph statistics and their standard deviation for Hypertext network\st{s} every $60$ minutes from June 2009 till July 2009. There are 48 graphs in total.}
\label{fig:hypertxt_40}
\end{figure}

\begin{figure}
    \centering
    \includegraphics[width=0.9\textwidth]{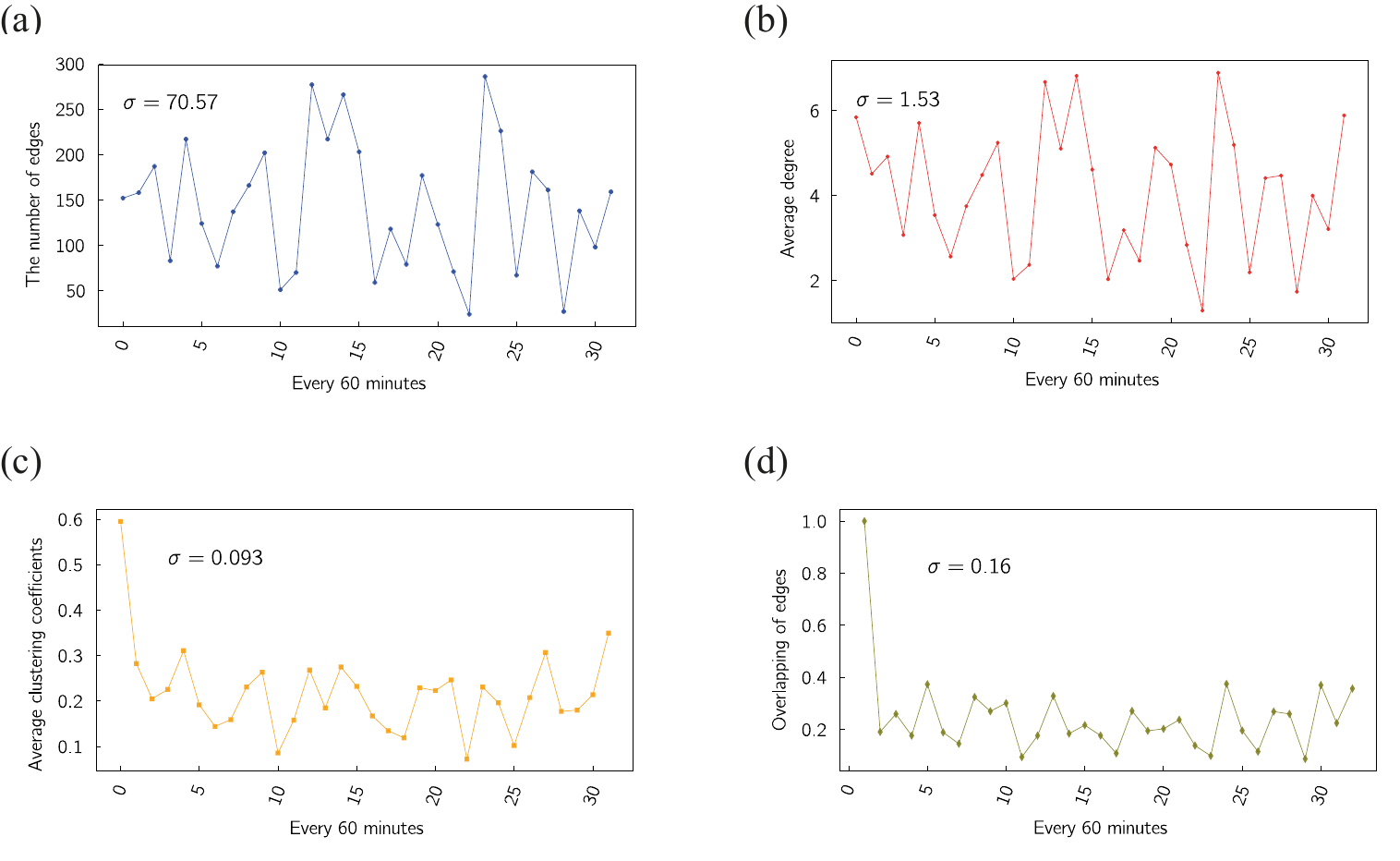}
    \caption{Temporal evolution of graph statistics and their standard deviation for Hypertext network\st{s} every $40$ minutes from June 2009 till July 2009. There are 32 graphs in total.}
    \label{fig:hypertext_60min}
\end{figure}

\begin{figure}
    \centering
    \includegraphics[width=0.9\textwidth]{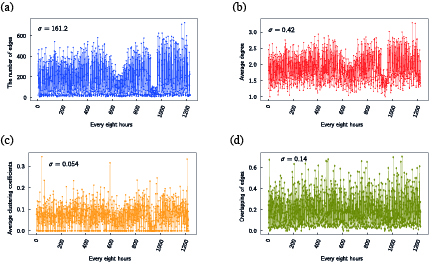}
    \caption{Temporal evolution of graph statistics and their standard deviation for \hl{E}mail network\st{s} every eight hours from Jan 1970 till March 1972.}
    \label{fig:email8h}
\end{figure}

\begin{figure}
    \centering
    \includegraphics[width=0.9\textwidth]{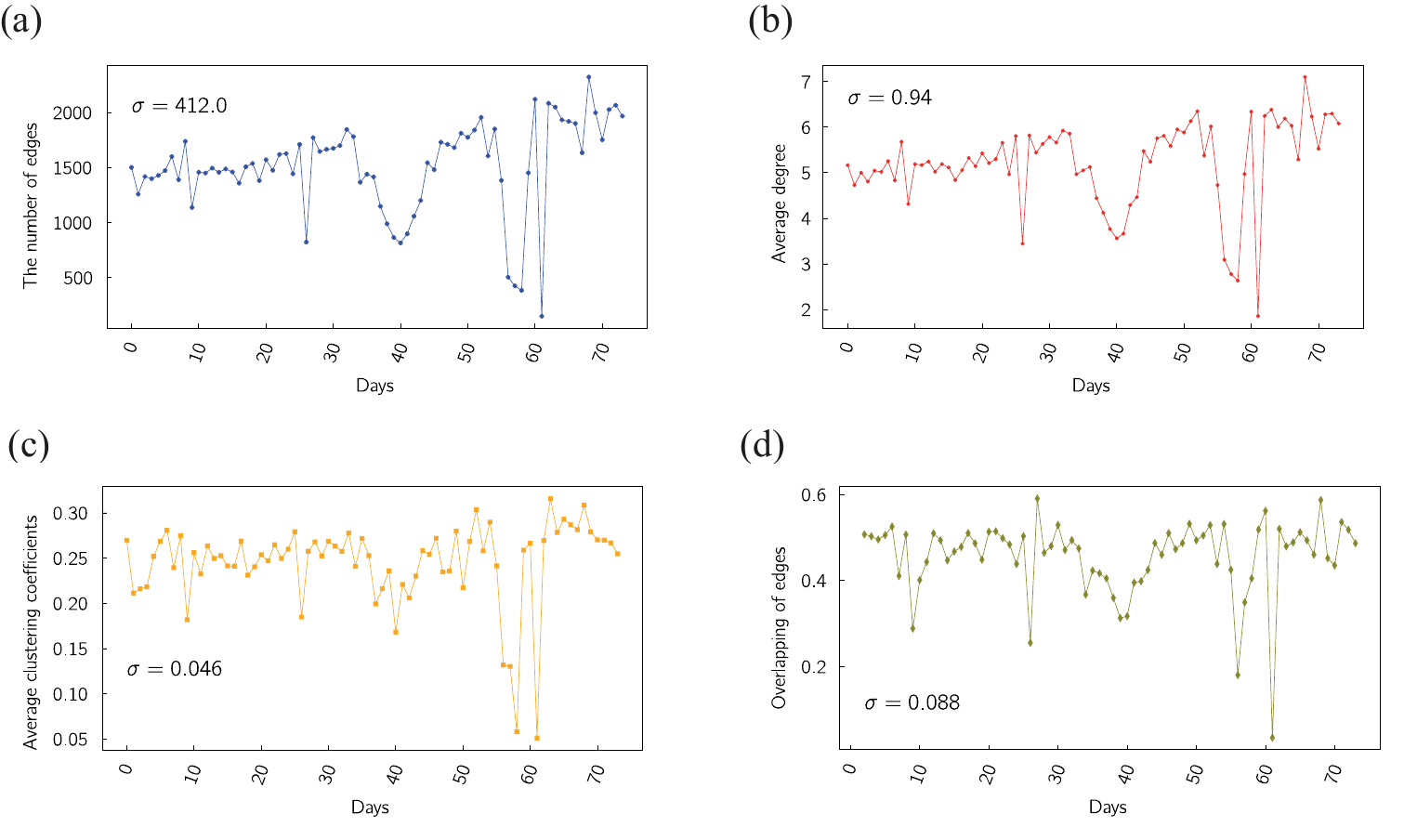}
    \caption{Temporal evolution of graph statistics and their standard deviation for \hl{E}mail \hl{N}etworks every seven days from Jan 1970 till March 1972.}
    \label{fig:email7days}
\end{figure}

\begin{figure}
    \centering
    \includegraphics[width=0.9\textwidth]{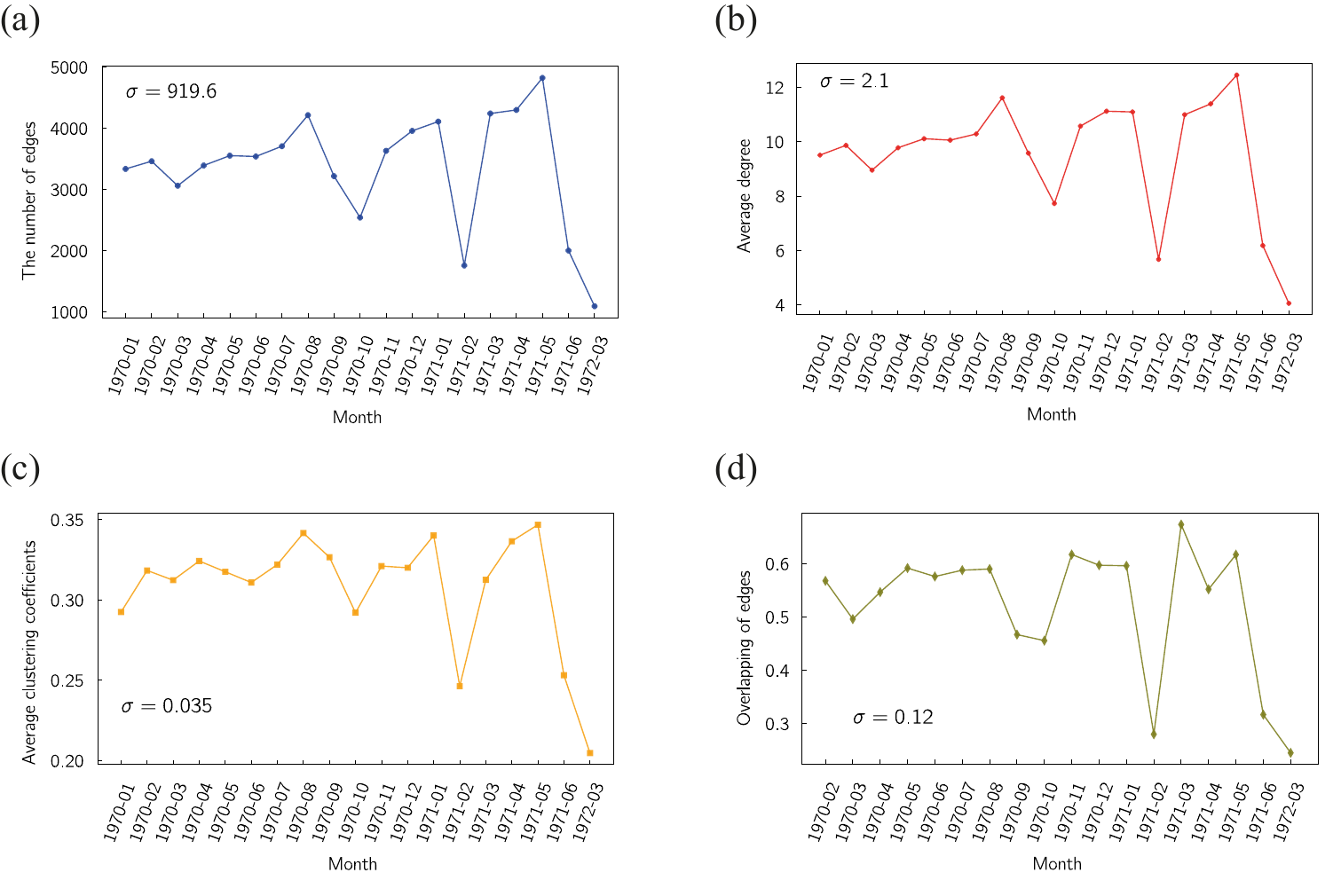}
    \caption{Temporal evolution of graph statistics and their standard deviation for \hl{E}mail network\st{s} every one month from Jan 1970 till March 1972. There are $19$ snapshots in total and the overlapping of edges starts from Feb 1970 that is the overlapping between Jan 1970 and Feb 1970.}
    \label{fig:email1month}
\end{figure}

\begin{figure}
    \centering
    \includegraphics[width=0.9\textwidth]{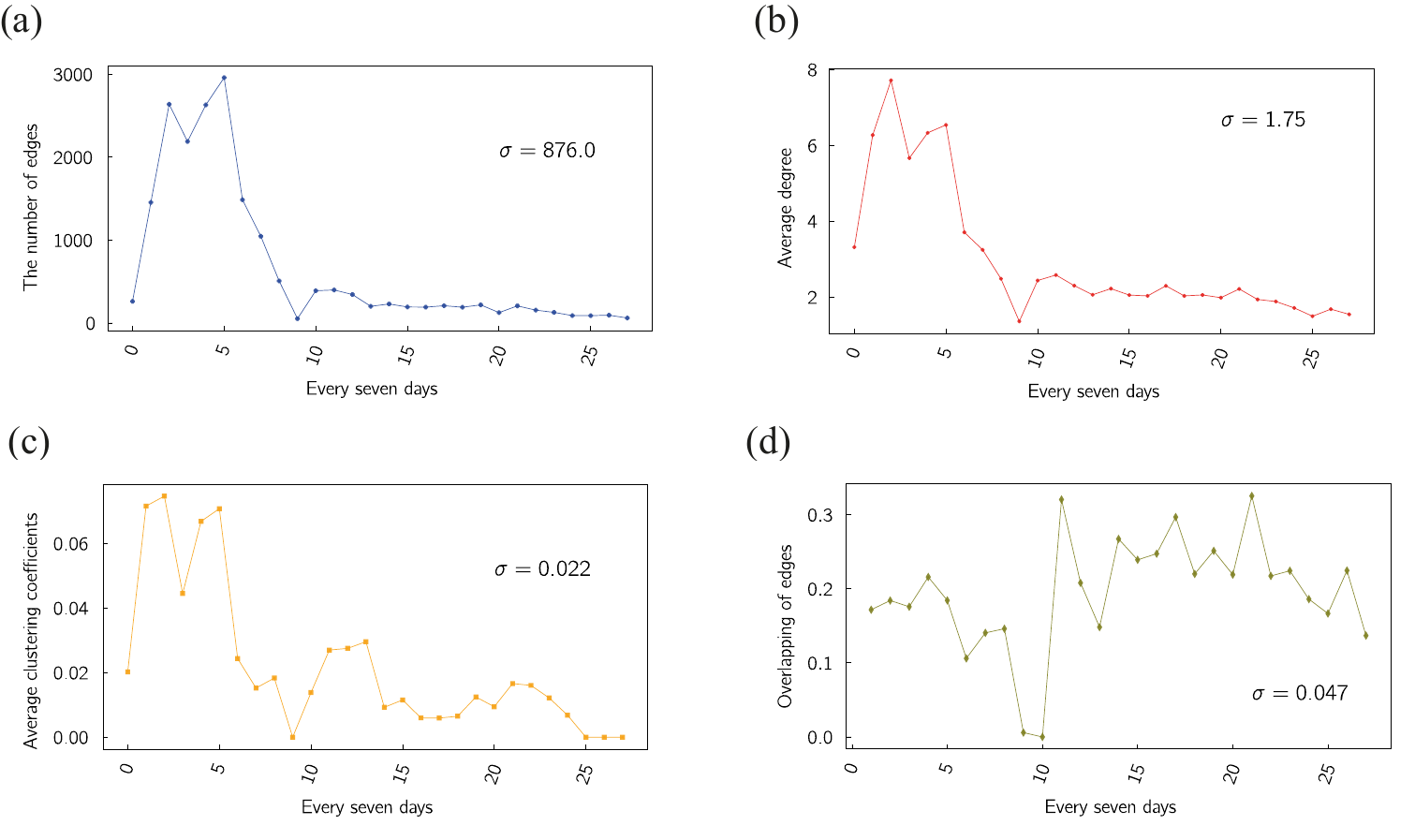}
    \caption{Temporal evolution of graph statistics and their standard deviation for College Message every 7 days from April 2004 till Oct 2004. There are $28$ snapshots in total.}
    \label{fig:college7day}
\end{figure}

\begin{figure}
    \centering
    \includegraphics[width=0.9\textwidth]{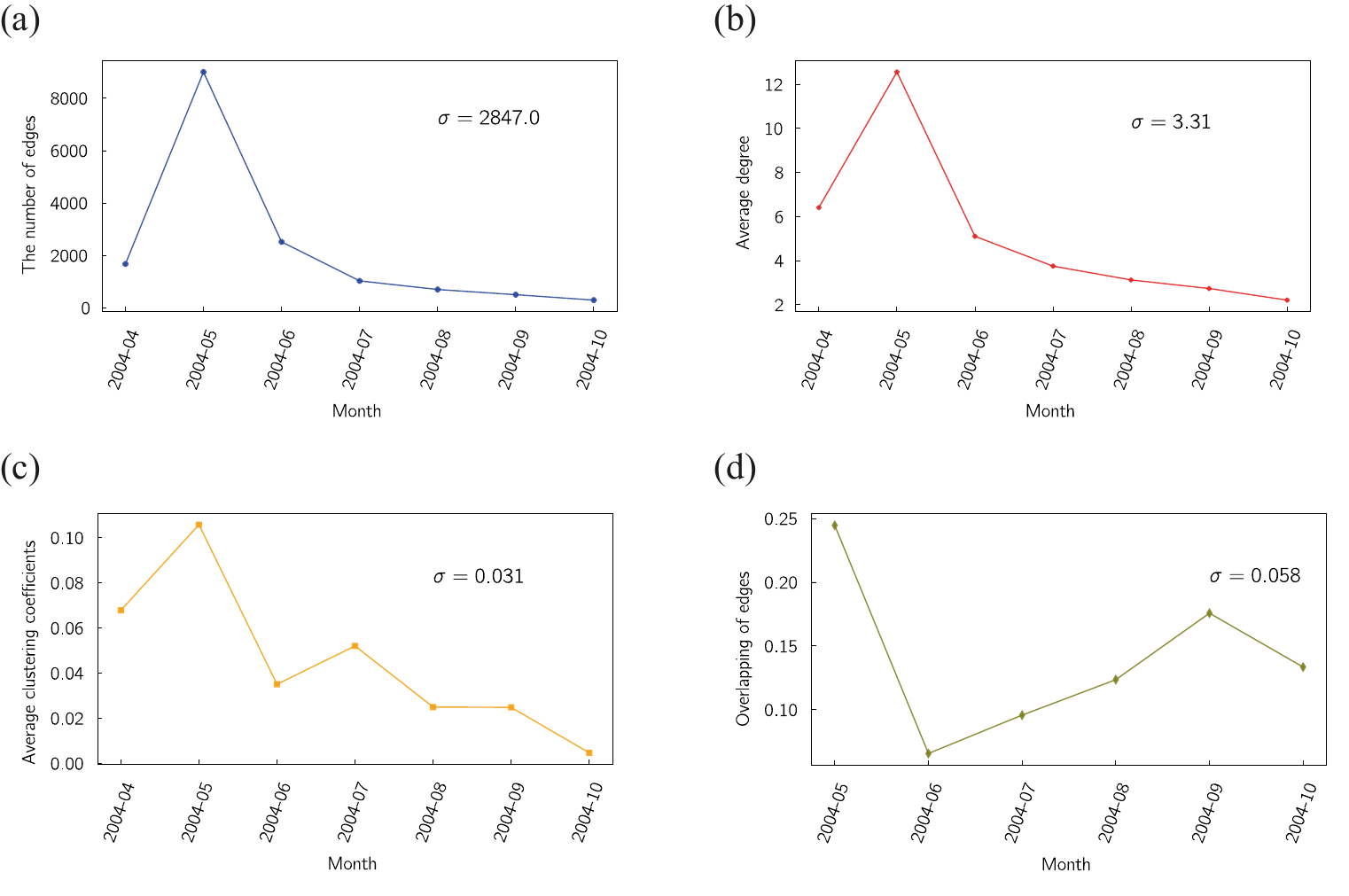}
    \caption{Temporal evolution of graph statistics and their standard deviation for College Message every month from April 2004 till Oct 2004. There are $7$ snapshots in total and the overlapping of edges starts from May 2004 that is the overlapping between April 2004 and May 2004.}
    \label{fig:college1mon}
\end{figure}
\clearpage
% ***********************************************************************************
\section{Significance test of the pairwise interactions}\label{a:zscore}

The significance of the transition of the triplet transition can be quantified by using the $z$-score (\href{https://en.wikipedia.org/wiki/Standard_score}{standard score}).  The $z$-score has been used to a qualitative measure of statistical significance of different motifs~\cite{MILO2002} or temporal motifs~\cite{KOV2013}. We apply similar procedures to compute $z$-scores for different triplet transitions and for a transition from graphlet $m_i$  to graphlet $m_j$ we define
\begin{equation}
    Z_{ij}= \frac{ \texpect{ \Tsfhat_{ij}(s) } - \texpect{ \Tsfhatpw_{ij}(s) } } { \sigma^{\mathrm{(pw)}}_{ij} } \, .
    \label{ae:Zdef}
\end{equation}
Here $\sigma^{\mathrm{(pw)}}_{ij}$ is the standard deviation in the $ij$-th entry of the transition matrices $\Tsfhatpw(s)$ obtained from the simple pairwise model.

To calculate this, we generate an ensemble of $R=1000$ realisations of the null model, the pairwise null model. The number of simulations $R$ needed was found as follows. We start from $R=100$ realisations, increase to $200$, $300$ and so on. Each time we increase $R$ we compare the difference in the results to those found with the $(R-100)$ realisations; if the results of the transition counting do not change from $(R-100)$ to $R$, we assume $R$ is sufficient, and we stop increasing $R$. Otherwise, we continue to increase the number of realisations until the results do not change.
The values of $p$ and $q$ used in the calculation of $\Tsfhatpw_{ij}(s)$ are those inferred
from one real network as described in  (6)
% ~\eqref{ae:phatdef}
and  (5).
% ~\eqref{ae:qhatdef}.

\begin{figure}
    \centering
    \includegraphics[width=0.8\textwidth]{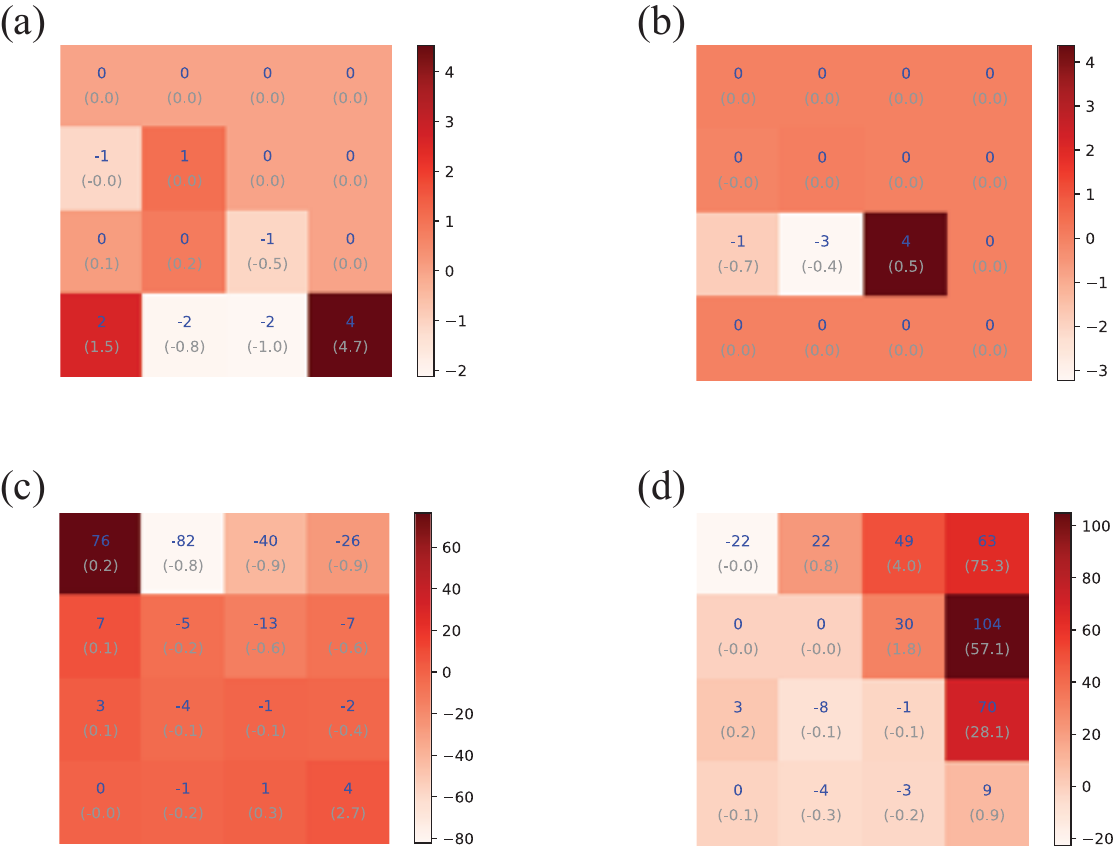}
    \caption{Z-scores for triplet transition based on comparisons of data to the pairwise simple  model.
    Each of the large four squares corresponds to a $Z$ matrix  \eqref{ae:Zdef} for a different data source, labelled as follows:
    (a) Turkish \hl{S}hareholder network ($\Delta t = 2 \mathrm{yr}$),
    (b) \hl{Wikipedia} Mathematician\st{s} ($\Delta t = 1 \mathrm{yr}$),
    (c) College \hl{M}essage data ($\Delta t = 1 \mathrm{mo}$), and
    (d) \st{Europe institutional e}\hl{E}mail network data ($\Delta t = 1 \mathrm{mo}$).
    %%%(a)Turkish shareholder network, $\Delta t$ is $2$ years; (b) Mathematicians wiki pages, $\Delta t$ is $1$ year; (c) College message data, $\Delta t$ is $1$ month; (d) Europe institutional email network data, $\Delta t$ is $1$ month.
    Each large square represents a four by four grid of smaller squares. The rows (columns) are the triplets at the earlier (later) arranged in order of size from smallest $m_0$ to largest $m_3$ going top to bottom (left to right). In each the colour represents $Z_{ij}$ on the scale given on the right of each large square, the top blue digits represent $Z$-scores,
    %\st{the middle black digits in brackets present standard deviation},
    and the bottom grey digits in brackets gives $( N^{\textsuperscript{r}}(m) - N^{\textsuperscript{pw}} ) / {N^{\textsuperscript{pw}}}$
    where $N^{\textsuperscript{r}}$ gives the number of triplets in the real data and $N^{\textsuperscript{pw}}$ gives the number predicted in the simple pairwise model for the same sized sample.}
    \label{fig:zscore_real}
\end{figure}

%\tnote{Can we find an analytic expression for $\sigma$? If so give these results, perhaps in an appendix? If we can, but we choose not to, then make a note of the reasons why (quicker and easier to use numerical estimate?)}\qcomment{This is a sensible idea, we need to do some literature reviews to check how to do it.}
A $z$-score $Z_{ij}$ with absolute value much bigger than one shows us that the data has behaviour not accounted for in our simple pairwise model. So it quantifies how likely a specific transition from state $i$ to $j$ is derived from higher-order processes not captured by simple pairwise interactions.
%%%The $\texpect{ \Tsfhatpw_{ij}(s) }_{ij}$ and $\sigma^{\mathrm{(pw)}}_{ij}$ are the average and the standard deviation of the counts of triplet transition from state $i$ to state $j$ of a sufficiently large number of an ensemble of pairwise models, respectively.
The results for $Z_{ij}$ for some of our networks are shown in \figref{fig:zscore_real}.

%\begin{equation}
%    Z_{ij}= \frac{\mathsf{C}_{r, ij}-\overline{\mathsf{C}}_{pw, ij}}{\sigma_{pw, ij}},
%\end{equation}

\clearpage
% *************************************************************************
\section{Other Link Prediction Methods}\label{a:othermethods}

%%%\tcomment{I feel we must define the measures we use in our paper, even if they are using definitions given elsewhere.  For that reasons I have reinstated some of the appendix on other methods. I am very happy for these to go in some online Supplementary Information file or, as here, in some appendix.}

We have used several other measures as summarised in main text and in \tabref{at:linkprediction}. These methods use node similarity measures to make link predictions and we will describe these scores in what follows.

%%%Here we give further details of methods: the \tdef{Common Neighbours} (CN), the \tdef{Jaccard Coefficient} (JC), \tdef{Resource Allocating Index} (RAI), the \tdef{Adamic-Adar Index} (AAI), the \tdef{Katz index} (Katz) method and the\tdef{Matrix Forest Index} (MFI).
% ............................................................
%\subsection{Similarity Measures}\label{a:similarity}
% In the following $s(u,v)$ is a (similarity) score assigned to a pair of vertices $u,v \in\Vcal$ where $\Ncal(u)$ is the set of neighbours of vertex $u$, i.e.\ $\Ncal(u) = \{ w | (u,w) \in \Ecal\}$.  The similarity scores are used to assign a probability $p(u,v)$ that an edge should exist between $u$ and $v$.

\begin{table}[htb!]
 \begin{center}
 \begin{tabular}{c||l| c| c| c}
 Abbreviation & Method                        & Reference    & Length Scale & Code\\ \hline\hline
 AAI  & Adar-Academic Index                   & \cite{LK07}  & $2$          & \texttt{nx} \\
 CN	  & Common Neighbour                      & \cite{LK07}  & $2$          & \texttt{nx} \\
 JC	  & Jaccard Coefficient                   & \cite{LK07}  & $2$          & \texttt{nx} \\
 Katz & Katz                                  & \cite{LJZ09a}& $\infty$     & \texttt{Own}  \\
 LLHN & Local Leicht-Holme-Newman             & \cite{LHN06} & $2$          & \texttt{Own} \\
 LPI  & Local Path Index                      & \cite{ZLZ09} & $3$          & \texttt{Own} \\
 EE	  & Edge Existence                        & [Here]       & $1$          & -   \\
 PA	  & Preferential Attachment               & \cite{LK07}  & $2$          & \texttt{nx} \\
 RA   & Resource Allocating Index             & \cite{ZLZ09} & $2$          & \texttt{nx} \\
 MFI  & Matrix Forest Index                   & \cite{CS97}  & $\infty$     & \texttt{Own} \\
 TT	  & Triplet Transition                    & [Here]       & $\infty$     & \texttt{Own} \\
 \end{tabular}
 \end{center}
 \caption{Table of the link prediction methods used and their abbreviations. The length scale given indicates the longest path length involved in the method or equivalently the largest power of the adjacent matrix involved in the method. Under code \texttt{nx} indicates that a NetworkX \cite{Networkx} routine was used, \text{Own} indicates the authors' own code was used. The Edge Existence (EE) approach was not investigated numerically but was included for the sake of comparison.}\label{at:linkprediction}
\end{table}

As we described in the main text, the most obvious similarity measure is the \tdef{Edge Existence} (EE) Index. It is defined as:
\beq
 s_\mathrm{EE} (u,v) = A_{uv} = \sum_{e \in \Ecal} \delta_{e,(u,v)} \, .
 \label{ae:EEmethod}
\eeq

The \tdef{Common Neighbours} (CN) method \cite{LK07,LJZ09a,AMA20} simply scores the relationship between two nodes based on the number of neighbours they have in common
\beq
 s_\mathrm{CN}(u,v) = \sum_{w\in \Vcal} A_{uw} A_{wv} = | \Ncal(u) \cap \Ncal(v) | \, .
\label{ae:CNmethod}
\eeq
This will tend to give large scores if $u$ and/or $v$ have high degrees.

To illustrate this, we can estimate the number of common neighbours in a random graph with degree distribution $p(k)$.  Suppose the vertices $u$ and $v$ have degree $k_u$ and $k_v$ respectively. Then this means we are picking out a pair of stubs with probability $k_uk_v/(4E^2)$ if there are $E$ edges in the simple graph. The probability that a pair of stubs are connected to the same nearest neighbour node of degree $k$ is $(1/2) k(k-1) p_\mathrm{nn}(k)$ where $p_\mathrm{nn}(k)= k p(k)/\texpect{k}$ is the probability that the nearest neighbour has degree $k$.  This gives us that the number of nearest neighbours in common in a random graph may be estimated to be
\beq
 s_\mathrm{CN,rnd}(u,v) \approx \frac{k_u}{2E}\frac{k_v}{2E} \frac{(\texpect{k^3}-\texpect{k^2})}{\texpect{k}}
 \propto s_\mathrm{PA}(u,v) \, .
 \label{ae:ncnrndg}
\eeq

One way to take this bias in $s_\mathrm{CN}(u,v)$ towards higher degree nodes is to make the score comparison between the two. If we compare by looking at the difference between $s_\mathrm{CN}$ and its expected value in the configuration model we end up with
\beq
 s_\mathrm{CN,diff}(u,v) = s_\mathrm{CN}(u,v) - \beta' s_\mathrm{CN,rnd}(u,v)
 =
 \left(\sum_{w\in \Vcal} A_{uw} A_{wv} \right) - \beta \frac{k_u}{2E}\frac{k_v}{2E}
\eeq
where $\beta$ is just a network-dependent rescaling of the parameter $\beta$.
This expression is very similar to a term in a Modularity index used for vertex clustering.  On the other hand, if we look at the fractional difference, we could use a score
\beq
 s_\mathrm{CN,frac}(u,v) = \frac{s_\mathrm{CN}(u,v) }{s_\mathrm{CN,rnd}(u,v)}
 \propto
 \frac{1}{k_u k_v} \sum_{w\in \Vcal} A_{uw} A_{wv}
\eeq
and we'll see similar forms in other indices below.

% Another way to compensate for the expected dependence of $s_\mathrm{CN}$ on the degree of the nodes is to normalise by the total number of unique neighbours.  This gives us the \tdef{Jaccard Coefficient} (JC) method \cite{LK07,AMA20} based on the well known similarity measure \cite{SM83}
% in which the likelihood that two nodes are linked is equal to the number of neighbours they have in common relative to the total number of unique neighbours.
The \tdef{Jaccard Coefficient} (JC) method \cite{LK07,AMA20} based on the well known similarity measure \cite{SM83}
in which the likelihood that two nodes are linked is equal to the number of neighbours they have in common relative to the total number of unique neighbours.
\bea
 s_\mathrm{JC}(u,v)
 &=& \frac{|\Ncal(u) \cap \Ncal(v)|}{|\Ncal(u) \cup \Ncal(v)|}
 \label{ae:JCmethod}
 \\
 &=& \frac{|\Ncal(u) \cap \Ncal(v)|}{|\Ncal(u)+|\Ncal(v)|-2A_{uv} - |\Ncal(u) \cap \Ncal(v)|}
 \\
 &=&
 \frac{s_\mathrm{CN}(u,v)}{k_u+k_v -2s_\mathrm{EE}(u,v) - s_\mathrm{CN}(u,v)}
  \, .
 \label{ae:JCmethod2}
\eea
For instance in a random graph we estimate that
\beq
 s_\mathrm{JC,rnd}(u,v)
 \approx
 \frac{s_\mathrm{CN,rnd}(u,v)}{k_u+k_v -2A_{uv} -s_\mathrm{rnd}(u,v)}
\eeq
which for large $k_1,k_2 \sim O(K)$ gives  $s_\mathrm{JC,rnd}(u,v) \sim O(K)$ while $s_\mathrm{CN,rnd}(u,v) \sim O(K^2)$.

The \tdef{Resource Allocating Index} (RAI) method \cite{ZLZ09} and the \tdef{Adamic-Adar Index} (AAI) method \cite{LK07,AMA20} are both based on the idea that if two vertices $u$ and $v$ share some `features' $f$ .
For our simple graphs, the most important feature two nodes can share is an edge between them, while the second most important feature is sharing a common neighbour. The Resource Allocating Index method and the {Adamic-Adar Index} method ignore the presence or presence of a direct connection
% \footnote{This would be where the features $f$ are edges.}
between vertices $u$ and $v$ (as captured by $s_\mathrm{EE}$ of \eqref{ae:EEmethod}) and the features considered for these two similarity scores are the the common neighbours themselves $w \in \Ncal(u) \cap \Ncal(v)$.  One way to picture this is to remember that the adjacency list representation of a network, often used in practice numerically, is where for each node we record the list of neighbours. We can picture this adjacency list of neighbours $\Ncal(u)$ as the `words' on the `document' $u$ to connect with text analysis methods.  The frequency of this feature simply is the degree $|\Ncal(w)|$ of the neighbouring vertex $w$ as that is the number of times this feature will occur in the adjacency lists of other vertices. The {Resource Allocating Index} method and the \tdef{Adamic-Adar Index} (AAI) method differ in their choice of weighting function $W(w)$ used to weight the features when defining their similarity score.

 The \tdef{Katz index} (Katz) method \cite{LK07,LJZ09a,AMA20} counts the number of paths between each pair of vertices, where each path of length $\ell$ contributes a factor of $\beta^\ell$ to the score.
\beq
 s_\mathrm{Katz}(u,v) = ([\unitmat-\beta \Amat]^{-1})_{uv}
\label{ae:Katzmethod}
\eeq

%%%\tnote{These two terms are basically picking up contributions we have in our triplet method? Can we say that Katz is a particular parametrisation of our method?}

In a random graph the probability the two stubs from vertices $u$ and $v$ of degree $k_u$ and $k_v$ respectively are connected is simply $k_u k_v /(4E^2)$.  Comparing this to the estimate for the number of common neighbours in a random graph,  \eqref{ae:ncnrndg}, we estimate that in a random graph the Katz score is dominated by the existence of an edge if $\beta \ll (\texpect{k^3}-\texpect{k^2})/2$.
(here $\beta=0.01$).

% The \tdef{Local Leicht-Holme-Newman index} (LLHN) method \cite{LZ11} is based on the vertex similarity index of \cite{LHN06}, but while this gives a specific motivation for the form, it is in the end just a specific rescaling of the Katz index  \eqref{ae:Katzmethod}, specifically
% \bea
%  s_\mathrm{LLHN}(u,v)
%  &=&
%  \frac{s_\mathrm{Katz}(u,v)}{s_\mathrm{PA}(u,v)}
%  \label{ae:LLHNmethod}
%  \\
%  &=&
%  \frac{s_\mathrm{Katz}(u,v)}{|\Ncal(u)| \, |\Ncal(v)|}
%  =
%  \frac{s_\mathrm{Katz}(u,v)}{k_u k_v}
%  \label{ae:LLHNmethod1}
%  \\
%  &=&
%   \Dmat^{-1} \left(\unitmat - \beta \Amat\right)^{-1} \Dmat^{-1}
%  \label{ae:LLHNmethod2}
% \eea
% where $\Dmat$ is a diagonal matrix whose entries are equal to the degrees of the nodes, $D_{uv} = \delta_{u,v} |\Ncal(u)|$.
% The motivation for using this normalisation is that $|\Ncal(u)| \, |\Ncal(v)|$ is proportional to the number of neighbours expected in the configuration model, as shown in   \eqref{ae:ncnrndg}.
% So the Local Leicht-Holme-Newman index is the Katz score relative to the Katz score expected for the same pair of nodes in the configuration model.

The \tdef{Matrix Forest Index} (MFI) method \cite{CS97}
is defined as
\beq
 s_\mathrm{MFI}(u,v) = [( \unitmat + \Lmat )^{-1}]_{uv} \, ,
 \label{ae:MFImethod}
\eeq
where $\Lmat = \Dmat - \Amat$ is the Laplacian.
%%%One way to interpret this measure is that it is the ratio of the number of spanning rooted forests which are $u$ and $v$ are part of the same tree rooted at $u$ to all spanning rooted forests of the network\footnote{No idea what this means}.
One way to see this forms a suitable similarity measure is to know that if $\Qmat=( \unitmat + \Lmat )^{-1}$, then $D_{ij}=Q_{ii}+Q_{jj}+Q_{ij}+Q_{ji}$ is a metric on the network and hence is a good distance measure.  Similarity measures are often related to distance through a function that ensures the similarity measure increases if the distance between two vertices decreases.

A final way to view the {Matrix Forest Index} method is to consider a discrete-time version of the diffusion process described by a Laplacian.  If we have a vector $\wvec(t)$ whose entries represent the number of particles at every vertex at time $t$, then we can define a diffusion process where
\beq
 \wvec(t+1) - \wvec(t) = \lambda \Lmat \wvec(t) \, .
\eeq
In this process, a fraction $\lambda$ of the particles at each node flow down each edge in each time step (so $\lambda k$ particles leave each node at each step so large degree nodes to lose a larger fraction of every time step)).  The Laplacian gives the network flow into a given vertex, and the number of particles is conserved (since $\sum_i L_{ij} +0$). The matrix $\Qmat=( \unitmat - \Lmat )^{-1}$ used to give the MFI score is therefore
\beq
 \wvec(t) = [( \unitmat + \Lmat )^{-1}] \wvec(t+1)  \, .
\eeq
That is if we were to demand that at time $t+1$ we had only had particles at one site $u$, then $Q_{uv}$ tells us how many of those particles were at vertex $v$ at the previous time step.

\subsection{Scores for other link prediction methods from under-sampling   }

\begin{figure}
    \centering
    % Put this figure back in
    \includegraphics[scale=1]{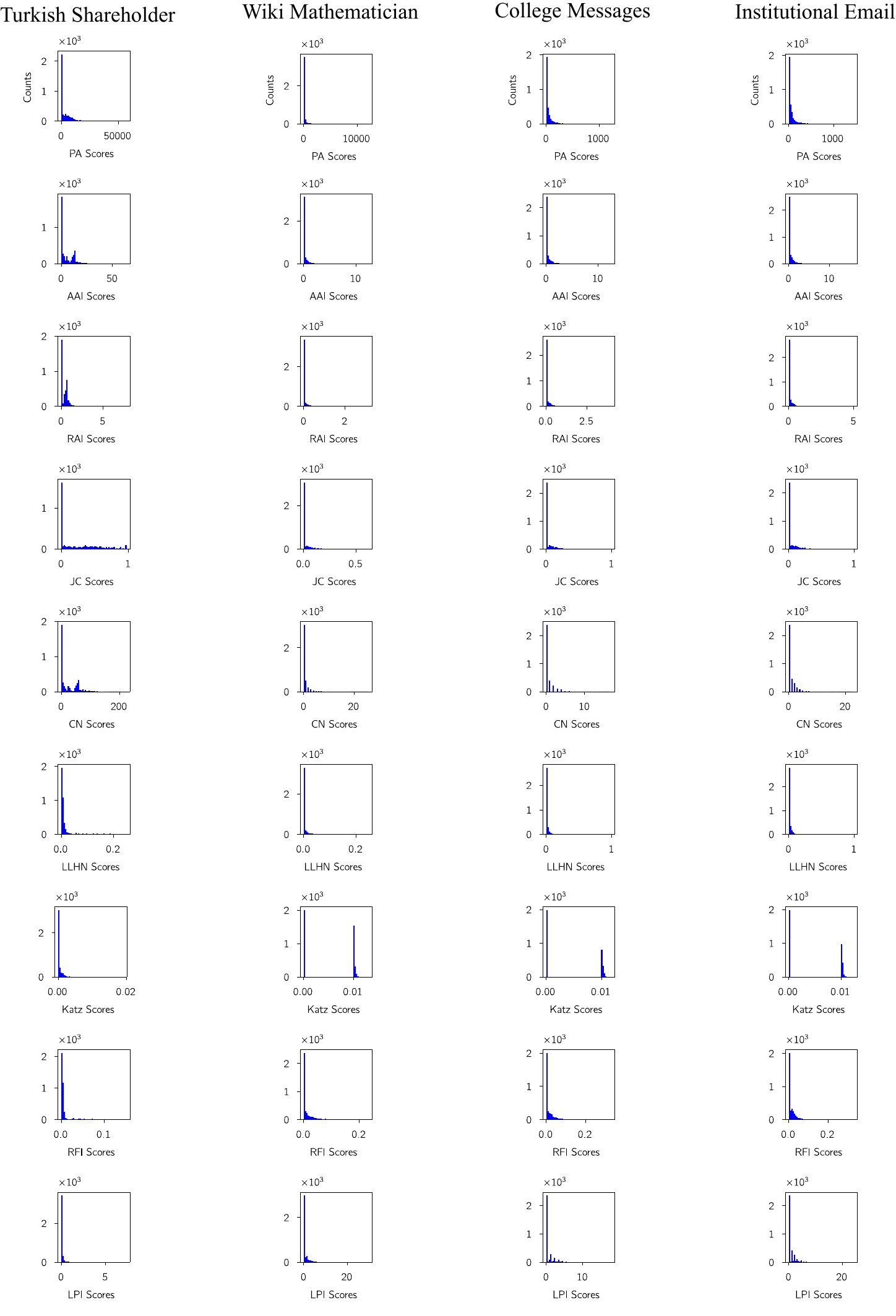}
    % \texttt{otherhist}
    \caption{Histograms of similarity scores for nine different link prediction methods applied to four different real world data sets.
    Each column is for a different data set, from left to right: Turkish \hl{S}hareholder \hl{N}etwork\st{s}, Wikip\hl{edia} Mathematician network\st{s}, College \hl{M}essage network, and Email network\st{s}.
    Each row is for a different link prediction method, see~\tabref{at:linkprediction} for abbreviations.
    The precise values are not important here as the important feature is the success or failure to identify two clear groups of low and high scoring node pairs.
    A successful method should have two separate peaks clearly visible in these plots and only the Katz method shows this on a regular basis.
    }
    \label{afig:other_hist}
\end{figure}

We under-sampled $1000$ link pairs and $1000$ non-connected pairs to demonstrate that TT can naturally split the score into two clusters\cite{BR03}, which can be used to predict link existence in main text Figure 4. We also compute the performance of other similarity measures, which shows, apart from Katz score, none of the methods show a good separation into two clear clusters in~\figref{afig:other_hist}.

\begin{figure}
    \centering
    % Put this figure back in
    \includegraphics[scale=1]{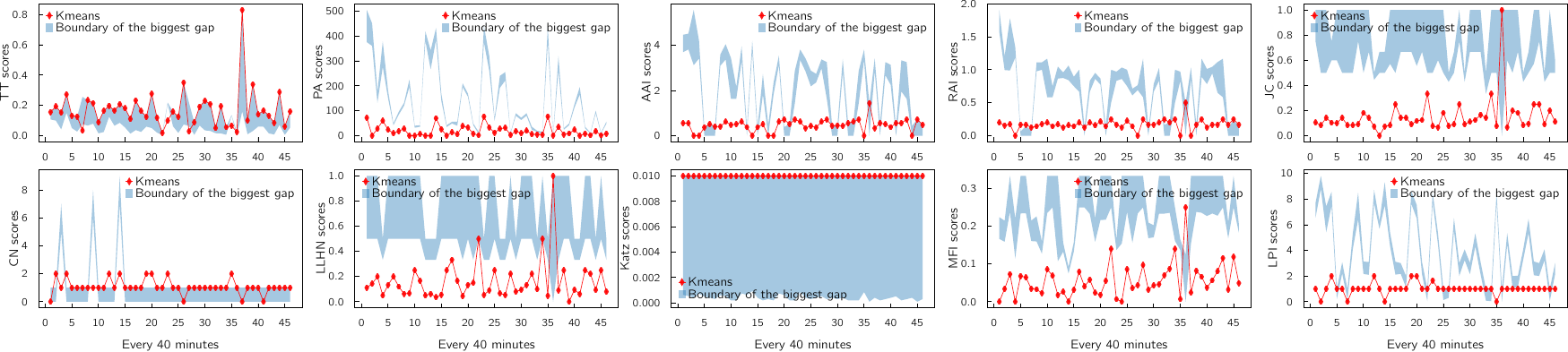}
    % \texttt{otherhist}
    \caption{The thresholds evolution plots for Hypertext network\hl{s} of $dt= 40$ minute. The red line represent the equivalent threshold of Kmeans clustering method which we use to do predictions and the blue filled space represents the boundaries of the biggest gap.}
    \label{afig:threshold_hyper_40}
\end{figure}

\begin{figure}
    \centering
    % Put this figure back in
    \includegraphics[scale=1]{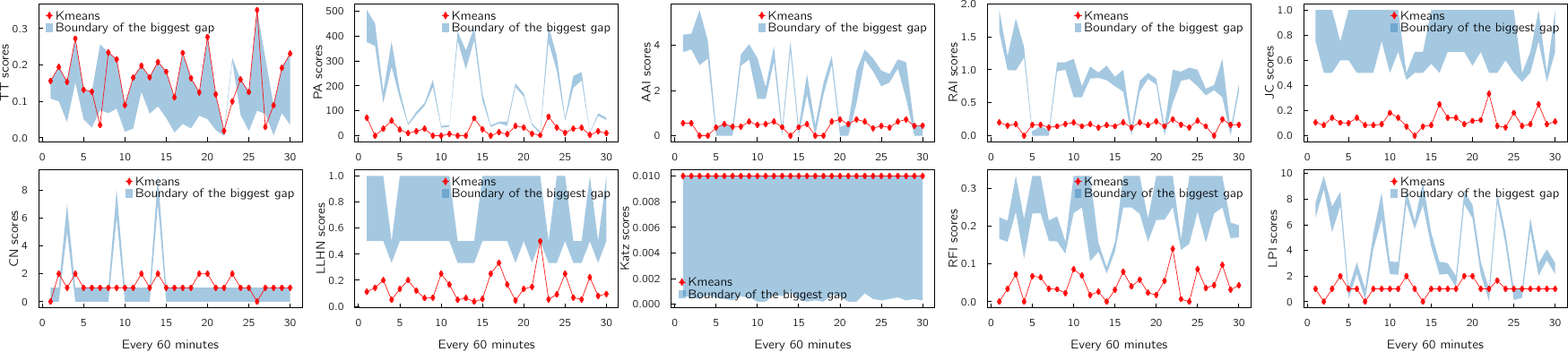}
    % \texttt{otherhist}
    \caption{The thresholds evolution plots for Hypertext network\st{s} of $dt= 60$ minute. The red line represent the equivalent threshold of Kmeans clustering method which we use to do predictions and the blue filled space represents the boundaries of the biggest gap.}
    \label{afig:threshold_hyper_60}
\end{figure}

% ...................................................
\subsection{From Node Scores to Link Predictions}\label{s:simtopred}

Once we have a similarity score for a pair of nodes, we have to turn this into a prediction. Generally, the node pairs with a high similarity score in $\Gcal(s)$ will be predicted to have an edge in $\Gcal(s+1)$, and low scores will lead to a no edge prediction. If we are looking at a link addition problem \cite{LJZ09a,S00b,PU03,BNG07,LK07,YCYTX07,CMN08} or, more generally, an uncertain link problem \cite{AMA20}, then the score is turned into a prediction for the node pair by using a standard machine learning approach to what is a binary classification problem.  For instance, we remove edges (or add an edge between unconnected nodes \cite{AMA20}) using some examples (say 10\%) to train the classifier and then the remain node pairs as used to verify the effectiveness of the method. In the simplest method we could imagine ranking our node pairs based on similarity scores, and then the $n_1$ unconnected node pairs with the highest scores are assigned an edge, while the $n_0$ lowest-scoring node pairs which are connected are predicted to have their edges removed.  The values for $n_0$ and $n_1$ could be learnt as part of the training of this simple classifier.  However, more sophisticated methods are normally used.

%%%In \cite{LJZ09a} ``In some link prediction algorithms, the scores may be not directly related to a certain similarity measurement but de-
%%%scribe the existence likelihood of links'' refers then to
%%%\cite{S00b} R. R. Sarukkai, Comput. Netw. 33, 377 (2000).
%%%\cite{PU03} A. Popescul and L. Ungar, Workshop on Learning Statistical
%%%Models from Relational Data, ACM Press, New York, 2003.
%%%%%%\cite{ZHH02} J. Zhu, J. Hong, and J.-G. Hughes, Proceedings of the13th
%%%%%%ACM Conference on Hypertext and Hypermedia ?ACM Press,
%%%%%%New York, 2002.
%%%\cite{BNG07} M. Bilgic, G. M. Namata, and L. Getoor, Proceedings of the
%%%7th IEEE International Conference on Data Mining, Work-
%%%shops ?IEEE Press, Washington, D.C., 2007
%%%\cite{YCYTX07}  K. Yu, W. Chu, S. Yu, V. Tresp, and Z. Xu, Advance in Neural
%%%Information Processing Systems 19 ?MIT Press, Cambridge,
%%%MA, 2007?.
%%%\cite{CCN08} A. Clauset, C. Moore, and M. E. J. Newman, Nature ?London?
%%%453, 98 ?2008?.

% ************************************************************************
\section{Measures of Success and Baseline Scores}\label{as:measures}

The notation $N_{\alpha\beta\pm}$ is the number of node pairs that
\begin{itemize}
\item start in state $\alpha$ ($1$ if the node pair is connected by an edge, $0$ otherwise) in $\Gcal(s)$,
\item which change to state $\beta$ defined in the same way but in terms of the existence of an edge between the same node pair, in snapshot $\Gcal(s+1)$,
\item for which the prediction made for that node pair was correct ($+$) or incorrect ($-$).
\end{itemize}
So if the prediction for node pair $(i,j)$ from snapshot $s$ to snapshot $(s+1)$ is $P(i,j)$ then
\beq
 N_{\alpha\beta +} =  \sum_{(i,j)} \, \delta(\alpha , A_{ij}(s) ) \, \delta( \beta , A_{ij}(s+1) ) \, \delta ( P(i,j) , A_{ij}(s+1) )
 \label{ae:Ndef}
\eeq
where $\delta(a,b)=1$ if $a=b$ otherwise it is zero (the Kronecker delta function).
For later convenience we also define
\bea
 N_{\alpha\beta} &=& N_{\alpha\beta +} + N_{\alpha\beta -} =  \sum_{(i,j)} \delta(\alpha,A_{ij}(s)) \delta(\beta,A_{ij}(s+1))
 \\
 N_{\cdot \beta} &=& N_{0\beta} + N_{1\beta } =  \sum_{j} \delta(\beta,A_{ij}(s+1))
\eea

The precision score is the number of times we predict an edge to exist between node pairs in the later snapshot correctly (a true positive) divided by the number of times we predict an edge, correctly (true positive) or incorrectly (false positive)
\bea
 S_{\mathrm{prec}} &=& \frac{N_{11+}+ N_{01+}}{ N_{11+}+ N_{01+} + N_{11-}+ N_{01-} } \, .
 \label{ae:precdef}
\eea
A high precision score means we can trust that edges predicted by the algorithm will exist.

The recall is the fraction of positive identifications which were correct, so
\bea
 S_{\mathrm{rec}} &=& \frac{N_{11+}+ N_{01+}}{ N_{11+}+ N_{01+} + N_{10-}+ N_{00-} }  \, ,
 \label{ae:recdef}
\eea

The accuracy is the number of correct predictions (edge or no edge) in the final snapshot over the total number of predictions
\bea
 S_{\mathrm{acc}} &=&
 \frac{N_{11+}+ N_{01+}+N_{10+}+ N_{00+}}{ N_{11+}+ N_{01+}+N_{10+}+ N_{00+} + N_{11-}+ N_{01-}+N_{10-}+ N_{00-} }  \, ,
 \label{ae:accdef}
\eea

Our baseline model is that we predict that an edge will connect a node pair with probability $\rho(s)$ where this is the density of the network in snapshot $s$, that is the fraction of edge pairs with an edge in snapshot $s$ where for $N$ nodes in the network and ignoring self-loops ($i=j$ excluded)
\beq
 \rho(s) = \frac{\sum_{(i,j)} A_{i,j} }{\sum_{(i,j)} 1} = \frac{2}{N(N-1)} \sum_{(i,j)} A_{i,j}  \, .
\eeq
This baseline model is equivalent to just randomising the edges in snapshot $s$ as a prediction for the next snapshot; it doesn't even preserve the degree of the nodes.
This means that in our notation we have that
\beq
 N_{\alpha 1 +} =    \rho(s)  N_{\alpha 1 +} \, , \quad
 N_{\alpha 1 -} = (1-\rho(s)) N_{\alpha 1 -} \, , \quad
 N_{\alpha 0 +} = (1-\rho(s)) N_{\alpha 1 -} \, , \quad
 N_{\alpha 0 -} =    \rho(s)  N_{\alpha 0 -} \, .
\eeq

The baseline value of precision is
\bea
 S_{\mathrm{prec,base}}
 &=&
 \frac{\rho(s) N_{\cdot 1} }{ \rho(s) N_{\cdot 1} + (1-\rho(s)) N_{\cdot 1} }
 \\
 &=&
 \rho(s)  \, .
 \label{ae:precbase}
\eea
For recall we find the baseline value is
\bea
 S_{\mathrm{rec,base}}
 &=&
 \frac{\rho(s) N_{11} + \rho(s) N_{01+} }{ \rho(s) N_{11}+ \rho(s) N_{01} + \rho(s) N_{10} + \rho(s) N_{00} }
 =
 \frac{ N_{\cdot 1} }{ N_{\cdot 1} + N_{\cdot 0}  }
 \\
 &=&
 \rho(s+1) \, .
 \label{ae:recbase}
\eea
where $\rho(s+1)$ is the density of edges in the network in snapshot $(s+1)$.
Finally accuracy in our baseline model is
\bea
 S_{\mathrm{acc,base}}
 &=&
 \frac{\rho(s) N_{\cdot 1}+ (1-\rho(s)) N_{\cdot 0} }{ \rho(s) N_{\cdot 1} + (1-\rho(s)) N_{\cdot 0} + (1-\rho(s)) N_{\cdot 1} + \rho(s) N_{\cdot 0} }
 \\
 &=&
 \rho(s)  \rho(s+1) + (1-\rho(s))(1-\rho(s+1)) \, ,
 \label{ae:accbase}
\eea
where for definiteness we have written $\rho=\rho(s)$.

An even more naive model which gives us another reference value would be one in which we simply predict an edge for any given node pair $50\%$ of the time. In our notation, this is equivalent to saying that $N_{\alpha\beta+}=N_{\alpha\beta-}$.  In this case the precision  \eqref{ae:precdef} and accuracy  \eqref{ae:accdef} both equal one half while recall  \eqref{ae:recdef} is simply the fraction of connected node pairs in the network, the network density.

% ************************************************************************
\section{Hypertext Network}\label{as:hypertext}

For completeness, we give the Hypertext network results equivalent to those shown in Figures 3, 5 and 6 of the main text and in Figures 14 and D10 of the appendix.

\begin{figure}[!h]
    \centering
    \includegraphics[scale=1]{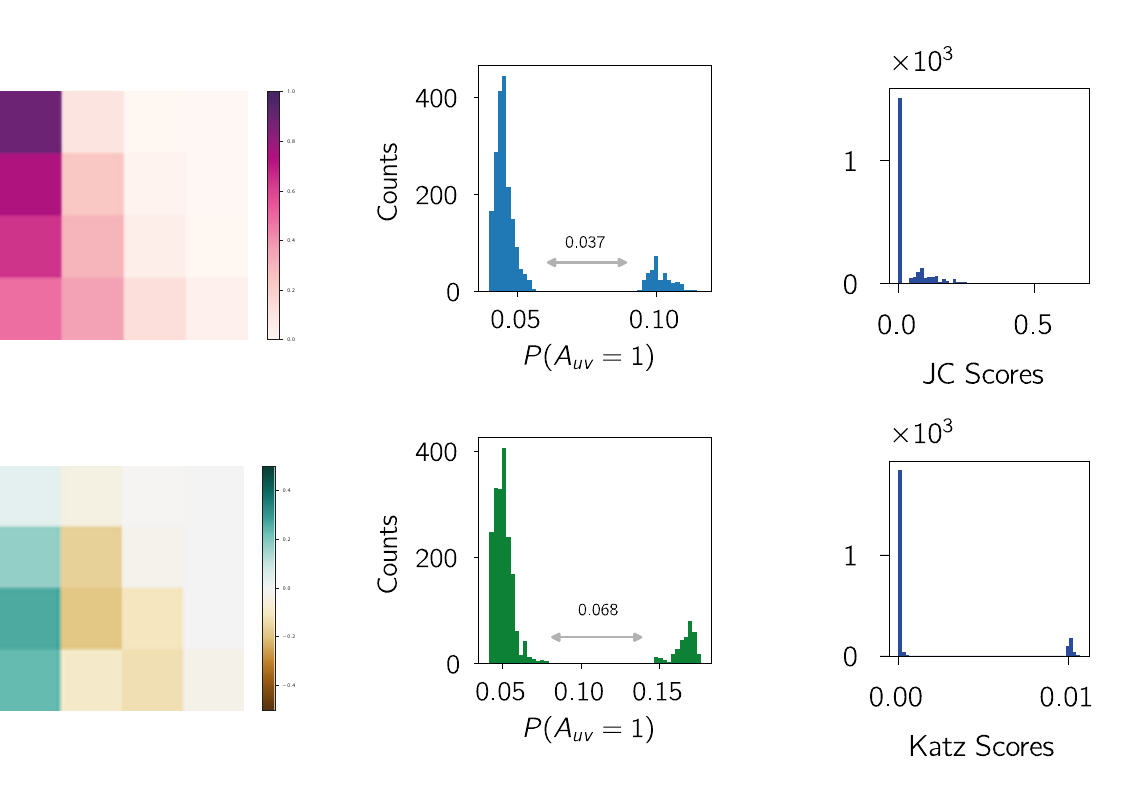}
    \caption{Results for the Hypertext network based using a  time window of 60min. Top left we show the transition matrix, bottom left we have a plot showing the entries of $\delta \That$. The central column shows the node similarity scores in our triplet transition (TT) method for the $\Mcal_4$ (top) and $\Mcal_8$ cases (bottom). The right hand column shows similar results for $\Mcal_4$ when using the Katz (top) and JC (bottom) methods.
    }
    \label{fig:hypertext_1}
\end{figure}

\begin{figure}[!h]
    \centering
    \includegraphics[scale=1.2]{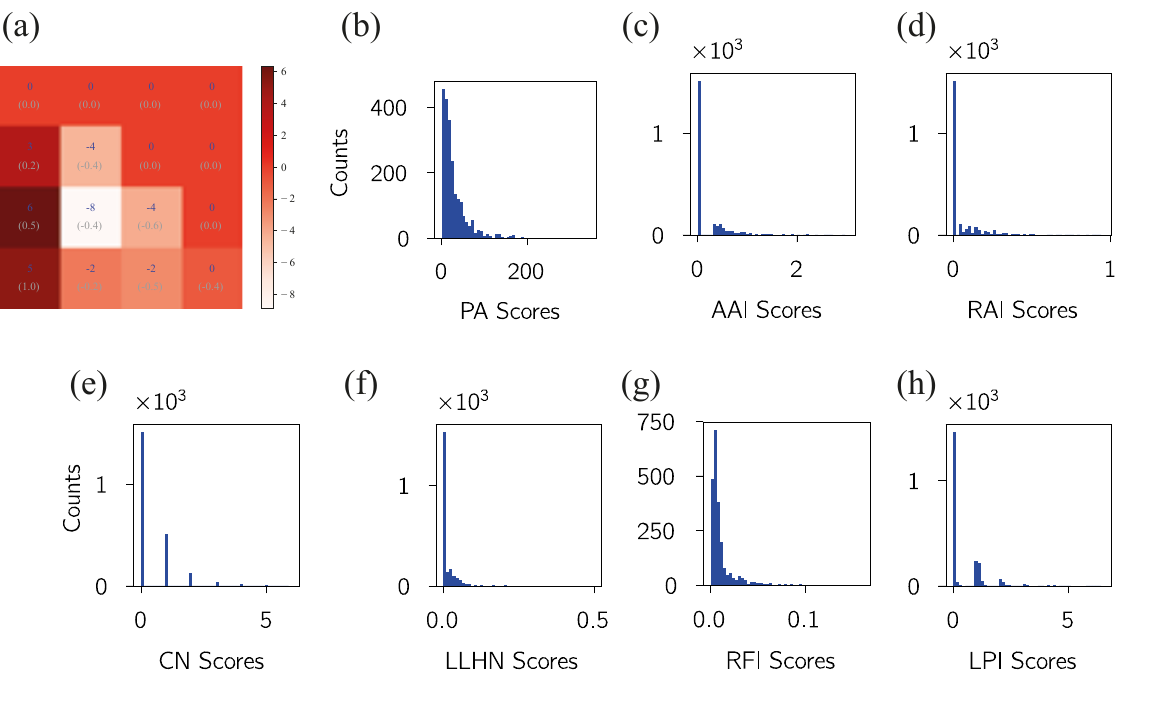}
    \caption{Results for the Hypertext network based using a  time window of 60min. Top left (a) we show the Zscore for transition matrix of Figure G14. (b)-(h) show the node similarity scores for other methods.}
    \label{fig:hypertext_2}
\end{figure}

\clearpage

\end{document}